\documentclass[preprint,12pt,authoryear]{elsarticle}

\usepackage{amsmath,amssymb,stmaryrd}
\usepackage{amsthm}
\usepackage[title]{appendix}
\usepackage{natbib}
    \usepackage[english]{babel}
    \usepackage[utf8]{inputenc}
\usepackage{color}

\DeclareMathOperator{\diff}{d}

\usepackage{mathtools}

\DeclarePairedDelimiter\abs{\lvert}{\rvert}

\newcommand{\xx}[1]{ [{{\color{red}\bf #1}}] }
\usepackage{algorithm2e}
\usepackage{caption}
\usepackage{subcaption}
\usepackage{algorithmic}

\usepackage{lipsum}

\usepackage{fancybox}

\usepackage{amssymb}

\usepackage{color}

\journal{Journal of Computational Physics}

\begin{document}

\begin{frontmatter}



\title{The synthesis of data from instrumented structures and physics-based models via Gaussian processes}


\author[1,2]{Alastair Gregory}
\author[1,2]{F. Din-Houn Lau}
\author[1,3]{Mark Girolami}
\author[1,4]{Liam J. Butler}
\author[5]{Mohammed Z. E. B. Elshafie}
\address[1]{Lloyd's Register Foundation's Programme for Data-Centric Engineering, Alan Turing Institute}
\address[2]{Department of Mathematics, Imperial College London}
\address[3]{Department of Engineering, University of Cambridge}
\address[4]{Lassonde School of Engineering, York University}
\address[5]{Civil and Architectural Engineering Department, College of Engineering, Qatar University}



\begin{abstract}

At the heart of structural engineering research is the use of data obtained from physical structures such as bridges, viaducts and buildings. 
These data can represent how the structure responds to various stimuli over time when in operation. Many models have been proposed in literature to represent such data, such as linear statistical models. Based upon these models, the health of the structure is reasoned about, e.g. through damage indices, changes in likelihood and statistical parameter estimates.
On the other hand, physics-based models are typically used when designing structures to predict how the structure will respond to operational stimuli. 
These models represent how the structure responds to stimuli under idealised conditions. What remains unclear in the literature is how to combine the observed data with information from the idealised physics-based model into a model that describes the responses of the operational structure.
This paper introduces a new approach which fuses together observed data from a physical structure during operation and information from a mathematical model. The observed data are combined with data simulated from the physics-based model using a multi-output Gaussian process formulation. The novelty of this method is how the information from observed data and the physics-based model is balanced to obtain a representative model of the structures response to stimuli.
We present our method using data obtained from a fibre-optic sensor network installed on experimental railway sleepers. The curvature of the sleeper at sensor and also non-sensor locations is modelled, guided by the mathematical representation. We discuss how this approach can be used to reason about changes in the structures behaviour over time using simulations and experimental data. The results show that the methodology can accurately detect such changes. They also indicate that the methodology can infer information about changes in the parameters within the physics-based model, including those governing components of the structure not measured directly by sensors such as the ballast foundation.

\end{abstract}




\end{frontmatter}



\section{Introduction}


The engineering research fields of structural health monitoring (SHM) and structural identification (SI) typically involve the collection and analysis of data from instrumented structures \citep{farrar2012structural,farrar2007introduction, ccatbacs2013structural, Aktan}. The primary objectives of SHM are to better understand the behaviour of structures (e.g. bridges, buildings, railway tracks, pipelines, tunnels etc.) over time and to identify and localise damage \citep{Cawley}. Novel sensing technologies such as fibre-optic and piezoelectric sensors have enabled the detailed collection of performance data from structures \citep{Sun}. These data are opening up the possibility for engineers to better reason about the structural condition of structures through statistical analysis (e.g. linear models \citep{Lau}).
Typical examples of response data collected from sensor networks on instrumented structures are in the form of vibration/acceleration, strain and acoustic responses \citep{Ye}. The data provides underlying information about how these structures respond to stimuli, and also includes noise due to various factors such as temperature effects and instrumentation error. Data is rarely measured across the whole domain of the structure either due to it being infeasible or too expensive \citep{Sazonov}. Therefore the response of the structure is only known at the sensor locations.


Most structures are designed and analyzed based on physical assumptions about how they should respond under various stimuli; these assumptions can be described using idealised mathematical models. These models typically are in the form of partial/ordinary differential equations \citep{Friswell, Sinha, Zeinali}. Partial differential equations (PDEs) often do not admit closed form solutions. Therefore they are often solved using e.g. finite element methods (FEM) \citep{Doebling}. Quantifying the uncertainty in the physically\footnote[4]{In the remainder of the paper we shall refer to the mathematical model of the physical system simply as the `physics-based model'.} modelled response of structures from PDEs, where there exists uncertainty in the input parameter, is a well-studied field in applied mathematics (e.g. for example data-assimilation, stochastic Galerkin methods, multilevel Monte Carlo \citep{Blondeel} and Markov chain Monte Carlo \citep{Dodwell}). There are several discrepancies between the observed data and the idealised physics-based model. First, the data represents the response of the sensor network not the structure: the underlying response of the structure is corrupted by the sensors. Second, the physics-based model is based on assumptions such as constant temperature and simplified structural properties. Further, the physics-based models do not take into account the instrumentation error (the error induced by the sensors). In the SHM literature it is not clear how to account for these discrepancies in a well-principled fashion that leads to a representative model of the data. 

Hybrid approaches are commonplace in literature, utilising data accrued from instrumented structures to improve predictions of structural response to stimuli from these physics-based models.
A commonly used example of a data/physics hybrid approach is model updating \citep{Rocchetta, Grafe, Schommer, Alkayem, Vigliotti}. In model updating, the parameters of the physics-based model are estimated using the measured response data; predicted responses of the structure are then given by the model with the estimated parameters. The parameters are typically estimated by optimising an objective function, such as the likelihood, which quantifies the discrepancy between the model and data \citep{Schommer}.
The predictive quality of the model updating approach relies on the physics-based model not incorporating too many simplifying assumptions and being closely representative of the actual response. In model updating, the physics-based model is treated as the true underlying data generating process \cite{Liu}. Here however, we assert that the observed data is the closest representation of the true structural response. Therefore the data should have an important role in the overall modelling procedure and prediction of structural response beyond estimating parameters.




In this paper, we introduce a novel modelling approach for structural response that balances the information from the observed data and the physics-based model. This will be referred to as a type of `data-centric engineering' (DCE) model, inline with the discussion in \cite{LauProb}. The information is obtained from the physics-based model by simulating data from it. These simulations are combined with observed data from the physical structure using a multi-output Gaussian process joint model \citep{OHagan, Raissi}. This model is constructed on the level of the geometric relationship between the physical quantities of the observed data and model simulations. Predictions of the structures response can be obtained via this joint model.
Unlike model updating, not only is observed data utilised to directly improve the physics-based model, but also the physics-based model is used to guide the inferred posterior from the data in unmeasured regions of the structural domain.
A multi-output Gaussian process model is also utilised in \cite{Zhou}; this differs from the work presented in this paper as it combines simulated data from two physics-based models with different fidelities (accuracies), instead of combining simulation from a physics-based model and observed data. Gaussian processes have also recently been used for other aspects of structural health monitoring in \cite{Neves, Worden, Teimouri, Fuentes}.

The benefits of the presented work are two-fold: First, through the proposed joint model one obtains aposteriori estimates for the response of the structure, inferred from both observed data and simulations from an analytical physics-based model. These two sources of information are balanced in order to improve the predictive performance of the posterior in regions where there is no measurement data available. Second, the modelling approach used estimates structural parameters within the physics-based model, even parameters governing components of the modelled system that cannot directly be measured by sensors. These capabilities make this work an important stride forward in the construction and evaluation of DCE models.


The response of railway sleeper beams (the components which carry the rail track and transfer the train axle forces into the supporting ballast and subgrade materials) will be used as a running example throughout this paper to demonstrate the proposed methodology. These sleepers can be modelled by the Euler-Bernoulli equation. This provides a simplified analytical physics-based model for the vertical deflection of the actual railway sleeper, although this can be generalised to richer models such as those that require the use of FEM. A core objective of SHM is to be able to reason about and detect structural change over time. The proposed multi-output Gaussian process model for the response of the railway sleeper can achieve this when used alongside change-point detection schemes. This is demonstrated in Sec. \ref{sec:results} using simulated and experimental data sets.


The paper proceeds as follows. Section \ref{sec:sensordata} concentrates on the data from sensor networks that is used throughout this work. The specifications of the instrumented railway sleeper used in the experimental aspect of this work is also discussed here. Following this in Sec.~\ref{sec:physics}, an analytical physics-based model for the vertical deflection of the sleeper is introduced. The main methodology of this paper is outlined throughout Sec.~\ref{sec:machinelearning}; this includes the estimation of the parameters within the physics-based model and the construction of a posterior for the response of the sleeper conditioned on data obtained from the sensor networks and physics-based model. This section also discusses by how much the physics-based model should inform the joint model, and proposes a principled approach to this problem. Finally in Sec.~\ref{sec:results}, simulated data and experimental data from an instrumented railway sleeper provide a demonstration on the effectiveness of the proposed methodology.


\section{Data from instrumented structure}
\label{sec:sensordata}

This study considers the modelling of a single horizontal prestressed concrete sleeper beam supported on compacted railway ballast. Data is obtained from the sleeper by instrumenting it with a network of fibre-optic sensors (FOS) consisting of Bragg gratings (FBG). FBG sensors interrogate light spectra and reflect them with a specific wavelength known as the `Bragg wavelength'; as the fibre optic cable containing the FBG is placed under strain at a given time $t_n$ this Bragg wavelength shifts linearly (from $\lambda_{t_0}$ to $\lambda_{t_n}$). The corresponding strain shift $\epsilon_{t_n}$ can then be computed via a linear transformation of this relative wavelength shift,
$$
\epsilon_{t_n} = 10^6 \left(\frac{\lambda_{t_n}-\lambda_{t_0}}{0.78 \lambda_{t_0}}\right).
$$
This strain is measured in units of microstrain, and measurements are recorded at discrete time indices $t_0,(t_0+\Delta t),\ldots,(t_0+n\Delta t)=t_n$ where $\Delta t=1/50$ seconds. Each FBG measures strain to an approximate accuracy of $\pm 4$ microstrain. A single fibre-optic cable can be inscribed with multiple FBGs. For a comprehensive review of these sensors see \cite{Kreuzer}, which also discusses how FBG sensors are affected by external factors such as temperature.

\begin{figure}[t!]
\centering
\begin{minipage}{\textwidth}
  \centering
  \vspace{18mm}
  \includegraphics[width=110mm]{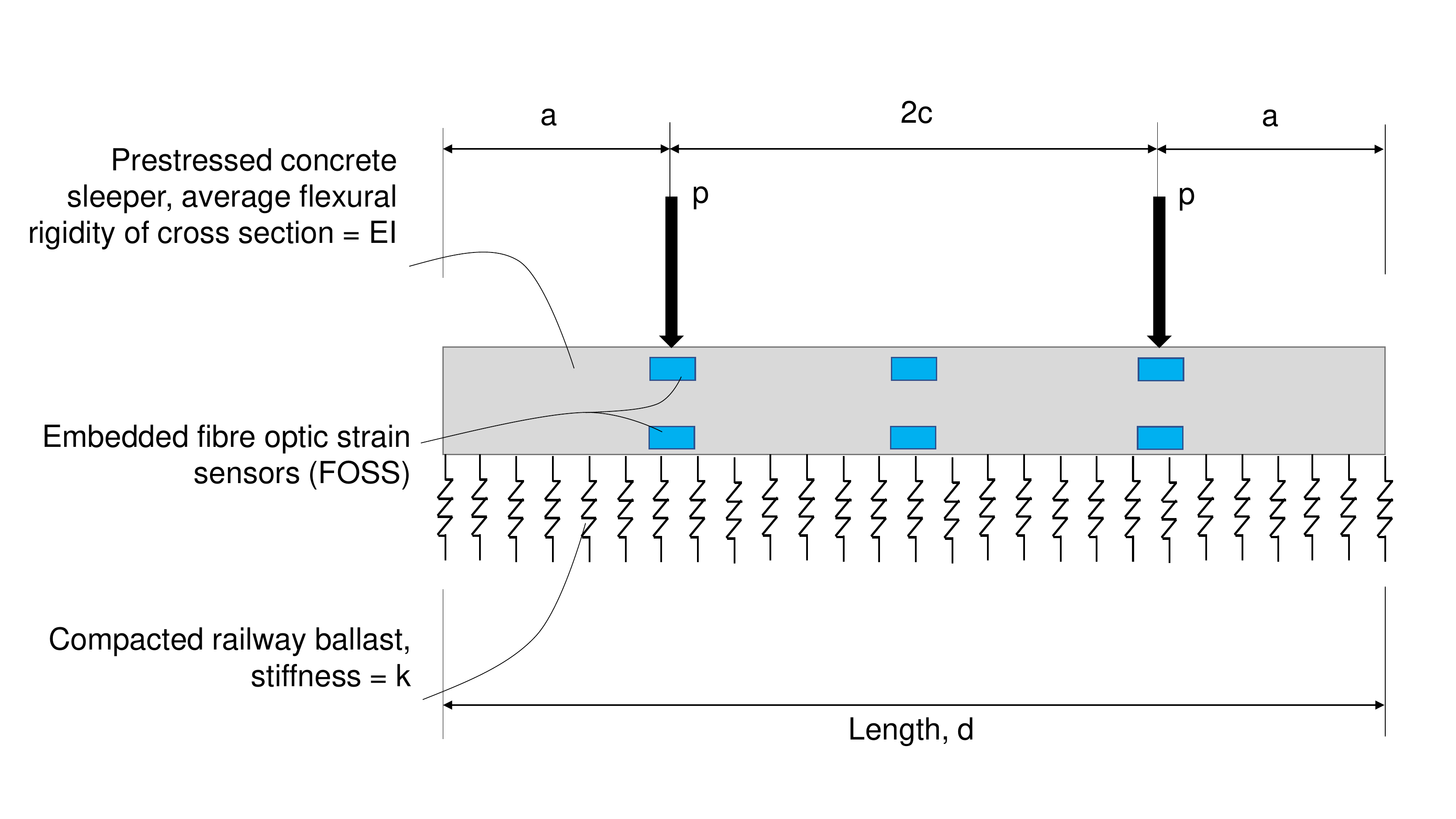}
\captionof{figure}{A schematic of the railway sleeper (supported on compacted ballast), instrumented with six fibre-optic sensors, with Bragg grating (FBG) considered in this study. These sensors are located on the top and bottom of the sleeper at the coordinates $x=x_1$, $x=d/2$ and $x=d-x_1$. The train wheels meet the sleeper at the two loading points, $x_1$ and $x_2$.}
\label{figure:schematic}
\end{minipage}%
\vspace{2mm}
\begin{minipage}{\textwidth}
  \centering
\includegraphics[width=100mm]{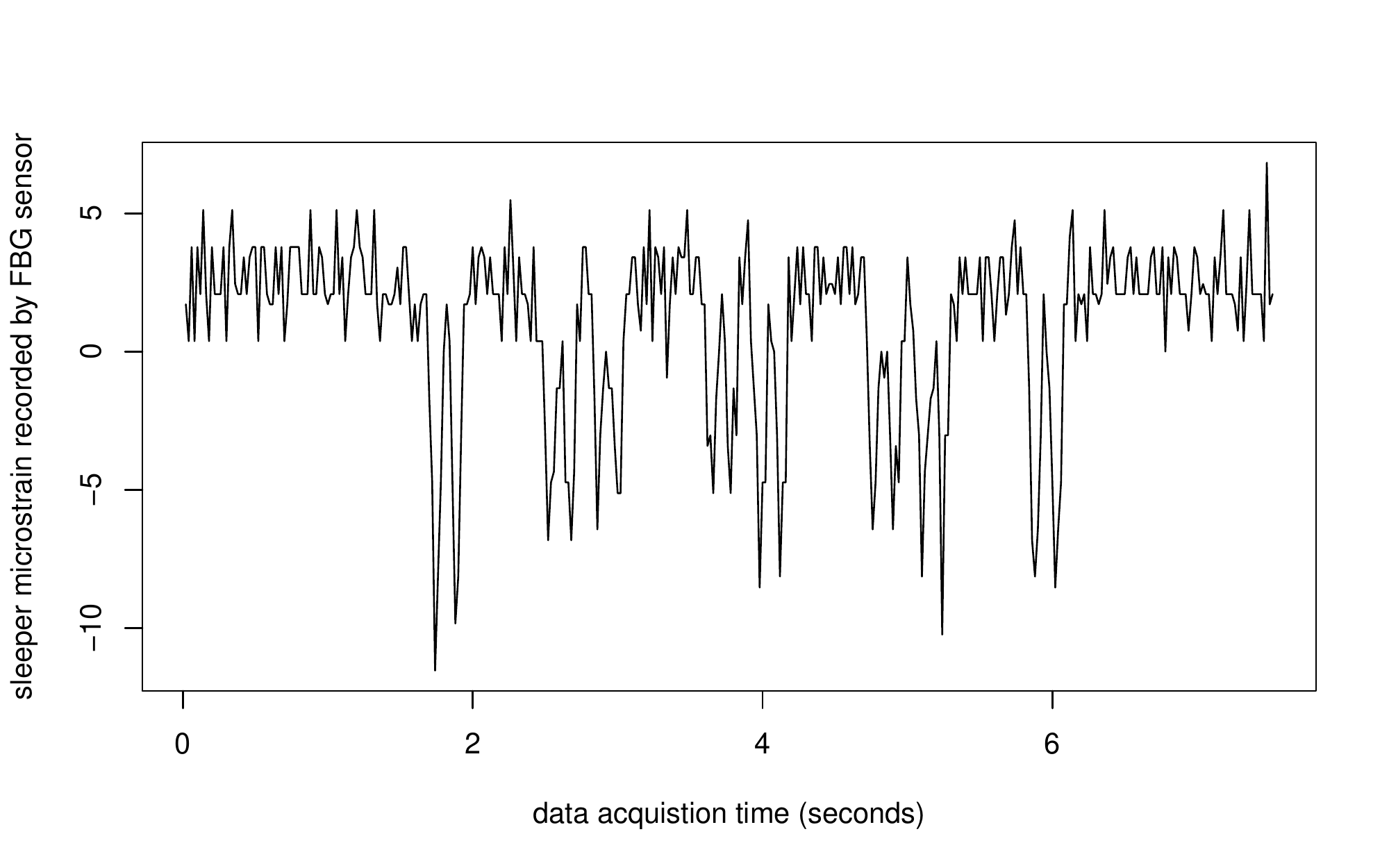}
\captionof{figure}{A 7.52 second sample of data from a fibre-optic sensor with Bragg grating acquiring data at 50Hz during the passage of a train. The sensor is located at $x=d-x_1$ along a sleeper.}
\label{figure:sensordata}
\end{minipage}
\end{figure}

The railway sleeper considered and used in the experimental results was instrumented during its fabrication with three FBG sensors attached along the top and bottom steel prestressing strands, which are embedded within the concrete sleeper. Coordinates on the sleeper are measured in millimeters. The sleeper is $d=2500$ by 200 (length/depth) with the FBG sensors located at $X_f=[X_{f_1},X_{f_2},X_{f_3}]=[500,1250,2000]$, on both the top and bottom prestressing strands embedded within the sleeper. The length of $X_f$ is denoted by $N_f$ for the benefit of notation later in the paper. In this particular application $N_f=3$. The distance between the top and bottom rows of FBG sensors on the instrumented sleeper is $91.5$. In experiments the sleeper was subjected to loading forces to simulate the axle spacing of the train wheels. Two equal forces were exerted by pistons at $x_1=(d-2c)/2$ and $x_2=d-x_1$, with $c=750$. The force magnitude was varied over the course of the experiment (see later Sec. \ref{sec:experiment} for details). 
A schematic of the sensor network instrumented on the sleeper beam is shown in Figure \ref{figure:schematic}. A more detailed specification of the instrumented sleepers considered in this study is given in \cite{Butler}. For illustrative purposes only, Figure \ref{figure:sensordata} shows a 7.52 second representative data set of strain measurements from an FBG (located at $x=2000$ along the top of an instrumented sleeper). This captures the moment when a train passes over the sleeper, resulting in the visibly large peaks of strain.

Given strain measurements $\epsilon_{t}(x)$ at the time index $t$ and one coordinate $x \in \mathcal{D}=[0,d]$ on both the top (denoted by the superscript $T$) and bottom (denoted by the superscript $B$) of the sleeper, the \textit{curvature} associated with these measurements can be computed by
\begin{equation}
z_t(x) = \frac{\epsilon^{(T)}_t(x)-\epsilon^{(B)}_t(x)}{91.5}.
\label{equation:sensorcurvature}
\end{equation}
Therefore the curvature here represents the gradient of strain over the depth of the sleeper. This combines strain measurements from the top and bottom of the sleeper into a single one-dimensional quantity $z_t(x)$. A one-dimensional domain simplifies the synthesis of this observed data with a one-dimensional physics-based model (presented in the next section). Curvature can be a useful tool in damage detection for beam structures \citep{Dawari}.


In practice, curvature data are only recorded at the sensor locations $X_f \in
\mathcal{D}$. Denote the curvature data at locations $X_f$ and time
index $t$ as $z_t(X_f) =
[z_t(X_{f_1}),z_t(X_{f_2}),z_t(X_{f_3})]^T$. We assume the $M$ curvature observations $\big\{z_i(X_f)\big\}_{i=1}^{M}$ are independent realisations from $z(X_f)$ and
\begin{equation}
  z(X_f) = \mu_f + W, \quad W \sim \mathcal{N}\left(0
    ,\Sigma_{f}\right),
\label{equation:noisemodel}
\end{equation}
where $\mu_f := f(X_f)$ is the true curvature and $\Sigma_f\in\mathbb{R}^{3\times 3}$ is the covariance matrix of
the Gaussian noise $W$. We assume herein that $\Sigma_f$ has a diagonal form, meaning that the noisy curvature observations at the different coordinates in $X_f$ are independent of one another.

We will obtain an alternative model for this curvature, conditioned on both the observed noisy data and information from a physics-based model in Sec. \ref{sec:machinelearning}. Curvature, $f(x)$, is geometrically related to other useful quantities for engineers, one being the vertical deflection, $y(x)$, of an idealized beam. These quantities are related via,
\begin{equation}
-\frac{\diff^2 y(x)}{\diff x^2}=f(x).
\label{equation:governingequation}
\end{equation}
This differential relationship is used in \cite{Xu} to estimate the deflection of beam structures from strain measurements. This relationship will be the basis of how curvature data from the instrumented sleeper will be synthesised with a physics-based model for vertical deflection in a joint model for the response of the sleeper under load (described in Sec. \ref{sec:machinelearning}). A physics-based model for the vertical deflection of the instrumented sleeper considered in this paper is described in the following section.

\section{Physics-based model description}
\label{sec:physics}

Engineers typically use physics-based models to assist in understanding how a structure behaves during excitation and at rest. These physics-based models are framed as solutions to (systems of) ordinary and partial differential equations, where often numerical discretization through finite element methods is required \citep{Doebling, Sinha}. This section describes an analytical physics-based model for the vertical deflection, $y(x)$, of a railway sleeper supported on compacted ballast \citep{Tran}.
The Euler-Bernoulli equation describes how a one-dimensional beam on the domain $\mathcal{D}$, with an elastic foundation, responds under a forcing $p(x)$. Such a response is utilised within the present work as a simplified representation of the response of the considered railway sleeper. The (static-time) Euler-Bernoulli equation is given by,
\begin{equation}
\frac{\diff^2 \left(EI(x)\left(\frac{\diff^2 y(x)}{\diff x^2}\right)\right)}{\diff x^2}=p(x),
\label{equation:governingEB}
\end{equation}
where the prime notation denotes the derivative with respect to $x$, $I(x)$ is the second moment of cross-sectional area and $E$ is the Young's modulus of the beam material (e.g. concrete). The product of $E$ and $I(x)$ is known as the flexural rigidity of the beam. We make the assumption that the flexural rigidity is constant over $x$, i.e. $I(x):=I\in (0,\infty)$.
See Sec. \ref{sec:conclusion} for a discussion on a possible way to relax this assumption.

\begin{figure}
  \centering
\includegraphics[width=100mm]{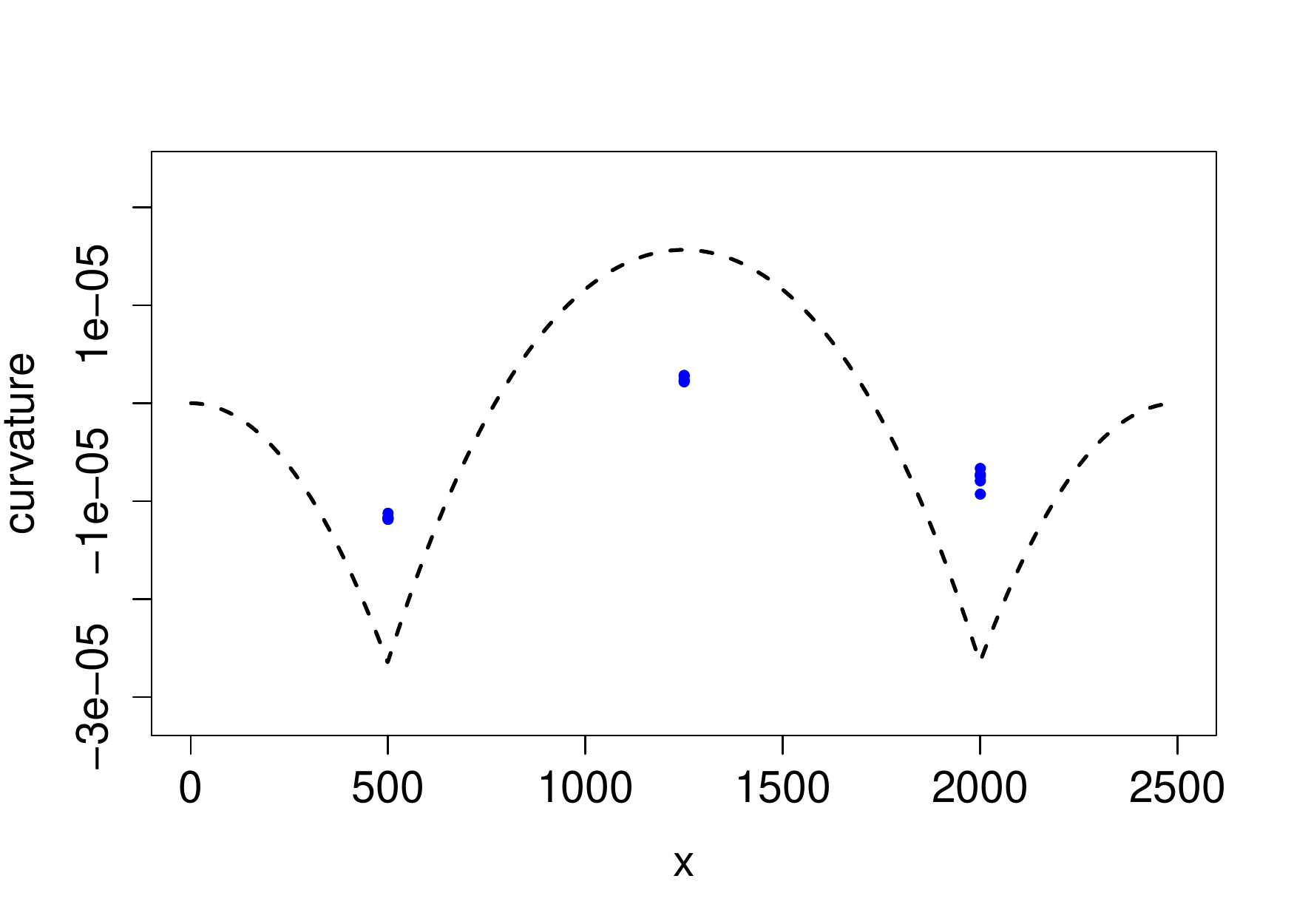}
\caption{An example 5 points of simulated curvature data $z_t(X_f)$ ($t=1,\dots,5$) at each coordinate $X_f$ along the sleeper. Also shown is $\mathcal{L}_{\phi,x}w(x,\lambda)$ numerically computed from the physics-based model of the vertical deflection $y(x)$ in a dashed black line. The parameters used in the model are given by prior beliefs in (\ref{equation:functionforregularization}). The forcing used is $p=125000$.}
\label{figure:data_and_physics}
\end{figure}

We obtain the physics-based model for the deflection of a railway sleeper by solving (\ref{equation:governingEB}), using a specific form for the forcing $p(x)$. In this case, an analytic solution is available.
Physics-based models do not necessarily describe how the physical system operates in practice, e.g. due to structural simplification such as assuming a constant flexural rigidity.
To derive the physics-based model for the deflection of a railway sleeper, we assume that during a train passing over the sleeper the forcing $p(x)$ has non-zero value at the two points $x_1$ and $x_2$ (experimentally simulated locations of train wheel axles), and $p(x_1)=p(x_2)=p$ (Newtons).
A schematic of the sleeper system (including the instrumented sensor network discussed in Sec. \ref{sec:sensordata}) is shown in Figure \ref{figure:schematic}. A solution to (\ref{equation:governingEB}) and an analytical model for the vertical deflection is \citep{Hetenyi},
\begin{equation}
y(x,p,k,\lambda) = \frac{p\lambda}{k}w(x,\lambda),
\label{equation:unitsolution}
\end{equation}
where
\begin{equation}
w(x,\lambda) =
\begin{cases}
v(x, \lambda) & x \in [0,x_1]\\
\begin{split}
v(x, \lambda) +\big[&\tilde{c}(\lambda (x-x_1))s(\lambda(x-x_1))+...\\
\quad &\tilde{s}(\lambda(x-x_1))c(\lambda(x-x_1))\big]
\end{split} & x \in [x_1, x_2],
\end{cases}
\label{equation:analyticdeflection}
\end{equation}
and
\begin{equation}
\begin{split}
v(x, \lambda) =\left[\tilde{s}(\lambda d)+ s(\lambda d)\right]^{-1}\Big\{&2\tilde{c}(\lambda x)c(\lambda x)\big[\tilde{c}(\lambda x_1)c(\lambda x_2)+\tilde{c}(\lambda x_2)c(\lambda x_1)\big]+...\\
\qquad &\big[\tilde{c}(\lambda x) s(\lambda x) - \tilde{s}(\lambda x)c(\lambda x)\big]\big[\tilde{c}(\lambda x_1)s(\lambda x_2) -...\\
\quad &\tilde{s}(\lambda x_1)c(\lambda x_2) + \tilde{c}(\lambda x_2)s(\lambda x_1) - \tilde{s}(\lambda x_2)c(\lambda x_1)\big]\Big\}.
\end{split}
\end{equation}
Here $c(\cdot)=\cos(\cdot)$, $s(\cdot)=\sin(\cdot)$, $\tilde{c}(\cdot)=\cosh(\cdot)$, $\tilde{s}(\cdot)=\sinh(\cdot)$. Finally $w(x,\lambda)=w(d-x,\lambda)$, for $x \in [x_2,d]$. Also $\lambda = \left(k/(4EI)\right)^{\frac{1}{4}}
$ is the flexibility of the sleeper and $k>0$ is the ballast stiffness. Let the parameters for this formulation of $y$, namely $p$, $k$ and $EI$ (which together lead to $\lambda$), be the components of the vector $\phi=(p,k,EI)$.
Using (\ref{equation:unitsolution}), one can rewrite the relationship between curvature and vertical deflection in (\ref{equation:governingequation}) as
\begin{equation}
\mathcal{L}_{x,\phi}w(x,\lambda)=f(x),
\label{equation:curvature}
\end{equation}
with the linear differential operator $\mathcal{L}_{x,\phi}=-\frac{p\lambda}{k}\frac{\diff^2}{\diff x^2}$. In Sec. \ref{sec:machinelearning}, we introduce a joint probabilistic model that uses (\ref{equation:curvature}) to combine observed curvature data (through strain data obtained from sensors at $X_f$; see Sec. \ref{sec:sensordata}) and the physics-based model in (\ref{equation:unitsolution}). Figure \ref{figure:data_and_physics} shows $\mathcal{L}_{x,\phi}w(x,\lambda)$ computed numerically (using finite differences) alongside an example 5 simulated curvature data points at each coordinate in $X_f$, where the values of $k=\exp(5)$ and $EI=\exp(28)$ used in $\phi$ are given by prior beliefs in (\ref{equation:functionforregularization}). The forcing used is $p=125000$.
In our method, we do not use the full analytic solution $w(x,\lambda)$ for all $x \in \mathcal{D}$; instead we simulate from $w(x,\lambda)$ at $X_u \in [0,d]^{N_u}$ and combine it with observed data in the joint model.
The only requirement here is that one is required to prescribe a value of $\lambda$ (estimation is used to circumvent this later in Sec. \ref{sec:parameterestimation}).
We use simulation from $w(x,\lambda)$ instead of using the full analytic solution (or infinitely many simulation coordinates) to create leverage on how (and where - unmeasured/measured areas of the domain) the physics-based model influences the joint model. If the full solution was used, the joint model would be fully informed by physics, and not by the observed data. In the case of a highly simplified model that poorly represents the observed response of the sleeper, this is detrimental.
On the other hand, in the case of a physics-based model that represents the observed response well, it would be desireable to incorporate many simulations into the joint model.
These examples inspire the process of `tuning' the choice of $X_u$; see Sec. \ref{sec:adaptivenu} for a discussion on this.


The analytical physics-based model presented in (\ref{equation:unitsolution}) represents a simplified response of the railway sleeper. It is used because it is inexpensive to simulate from and incorporates parameters of interest which govern the sleeper/rail trackbed system, including the ballast stiffness. In other cases, a sophisticated FEM model for $y(x)$ (or `digital twin') might be more appropriate; however, the number of times one can simulate from this may be limited by computational expense.
The next section describes how the physical simulations discussed in this section and the strain measurement-derived curvature data from the instrumented railway sleeper can be synthesised in a joint Gaussian process model.

\section{A data-centric engineering model based on Gaussian processes}
\label{sec:machinelearning}

This section will describe the methodology synthesising the physics-based model and the curvature data. The modelling of differential equations, such as 
\begin{equation*}\tag{\ref{equation:curvature} revisited}
\mathcal{L}_{x,\phi}w(x,\lambda)=f(x), 
\end{equation*}
is an important field of research for physical and engineering applications. In the case considered in this paper, it allows one to model the response of a railway sleeper under forcing by fusing curvature data and physical simulation for vertical deflection on the level of the geometric relationship between them. We will now describe the components of (\ref{equation:curvature}). First, recall that $f(x)$ represents the true curvature, which we do not have direct access to. Instead we observe the noisy curvature data $\big\{z_i(X_f)\big\}_{i=1}^{M}$ located at $X_f$ (see model \ref{equation:noisemodel}). We will describe later in this section how the observations are used to estimate the true curvature. Second, we define $u(x)=w(x,\lambda)$ for a prescribed $\lambda$; hence $u(x)$ represents the physics-based model for vertical deflection. A joint model for $u(x)$ and $f(x)$ using (\ref{equation:curvature}) is now described.

A recent landmark paper \cite{Raissi} proposes to learn about differential systems, such as (\ref{equation:curvature}), and the parameters within them by training multi-output joint Gaussian process models \citep{OHagan} based on observations from both $u(x)$ and $f(x)$. The multiple `outputs' in the model correspond to $u(x)$ and $f(x)$, respectively. This joint model is described presently.
Assuming a mean-zero Gaussian process prior for $u(x)$,
$$
u(x)|\sigma^2 \sim \mathcal{GP}(0, k_{u,u;\sigma^2}(x,x')) ,
$$
one can also construct a Gaussian process model for $f(x)$ via the differential operator $\mathcal{L}_{x,\phi}$. The infinite-dimensional stationary covariance kernel used for the prior is
$$
k_{u,u;\sigma^2}(x,x') = \exp \left\{-\frac{\sigma^2}{2}\left(\frac{x-x'}{d}\right)^2 \right\},
$$
where $\sigma^2$ is an unknown reciprocal length-scale parameter. This is known as the squared exponential covariance function and is a commonly used covariance kernel in many statistical applications given it's attractive properties, e.g. see \cite{Zhou}. For example, as $(x-x')^2$ increases from 0 at $x=x'$, the covariance between the two points decreases from it's maximum at 1 to approximately 0 when the two points are sufficiently `far away' from each other. This local dependence behaviour is found in many physical systems. Given the form of the covariance kernel, let $\theta=(\phi,\sigma^2)$ be the extended parameter set of the system. For $X=[X_1,\ldots, X_n] \in [0,d]^n$ and $X'=[X_1',\ldots, X_m'] \in [0,d]^m$, we define the matrix $k_{u,u;\sigma^2}(X,X') \in \mathbb{R}^{n \times m}$ as
$$
[k_{u,u;\sigma^2}(X,X')]_{i,j}= k_{u,u;\sigma^2}(X_i,X_j')\quad
i=1,\dots,n; j=1,\dots,m.
$$

The linear operator $\mathcal{L}_{x,\phi}$ can be used to obtain the covariance kernel of the Gaussian process that models $f(x)$, through
$$
f(x)|\theta = \mathcal{L}_{x, \phi} u(x)|\sigma^2 \sim \mathcal{GP}(0, k_{f,f;\theta}(x,x')),
$$
where
$$
k_{f,f;\theta}(x,x')=\mathcal{L}_{x,\phi}\mathcal{L}_{x',\phi}k_{u,u;\sigma^2}(x,x').
$$
Also define the cross-covariance kernels
$$
k_{u,f,\theta}(x,x') = \mathcal{L}_{x,\phi}k_{u,u;\sigma^2}(x,x'), \quad k_{f,u;\theta}(x,x')=\mathcal{L}_{x',\phi}k_{u,u;\sigma^2}(x,x').
$$
In this paper, we concentrate on the case where $\mathcal{L}_{x,\phi}=-(p\lambda\diff^2)/(k\diff x^2)$ from (\ref{equation:curvature}), and therefore the kernels $k_{u,f;\theta}$, $k_{f,u;\theta}$ and $k_{f,f;\theta}$ are given by
$$
k_{u,f;\theta}(x,x') = k_{f,u;\theta}(x,x')= -\frac{\sigma^2p\lambda}{d^2k}\left\{\left(\frac{\sigma(x-x')}{d}\right)^2 -1\right\}k_{u,u;\sigma^2}(x,x'),
$$
and
$$
k_{f,f;\theta}(x,x') = \frac{\sigma^4p^2\lambda^2}{d^4 k^2}\left\{3-6\left(\frac{\sigma(x-x')}{d}\right)^2+\left(\frac{\sigma(x-x')}{d}\right)^4\right\}k_{u,u;\sigma^2}(x,x').
$$
Given the physics-based model simulation coordinates $X_u$ and the
coordinates at which noisy strain measurement-derived curvature data
is available at $X_f$, one can write the joint distribution
$p(u(X_u), z(X_f)|X_u,X_f,\theta)$, where $z(X_f)$ is as defined in (\ref{equation:noisemodel}), as
\begin{equation}
\begin{bmatrix}
u(X_u)\\
z(X_f)
\end{bmatrix} \sim
\mathcal{N}\left(
\begin{bmatrix}
0\\
0
\end{bmatrix},
K_{\theta}
\right),
\label{equation:likelihood}
\end{equation}
where
$$
K_\theta=
\begin{bmatrix}
k_{u,u;\sigma^2}(X_u,X_u) & k_{u,f;\theta}(X_u,X_f)\\
k_{f,u;\theta}(X_f,X_u) & k_{f,f;\theta}(X_f,X_f) + \Sigma_f
\end{bmatrix},
$$
from \citep{OHagan, Raissi}. This model allows one to model the differential system in (\ref{equation:curvature}) probabilistically, estimate the parameters $\theta$ and construct point-wise posteriors for either $f(x)$ or $u(x)$. The next two sections describe how simulations from the physics-based model for vertical deflection $u(x)$ and noisy curvature data $\big\{z_i(X_f)\big\}_{i=1}^{M}$ can be used to estimate the parameters within the physics-based model and obtain aposteriori estimates for the sleeper response. The model in (\ref{equation:likelihood}) is a DCE model \citep{LauProb}, combining information about two different physical quantities, one from physical simulation and the other from measurement data obtained from instrumented structures.



\subsection{Parameter estimation}
In this section, we describe the procedure used to estimate the
parameters from the observation model in (\ref{equation:noisemodel}); $\Sigma_f$ and $\mu_f$ and the model in (\ref{equation:likelihood}); $\theta=(p,k,EI,\sigma^2)$. In the case where the forcing term $p$ is known, the parameter vector $\theta$ reduces to $(k,EI,\sigma^2)$. This is the case considered in the remainder of the paper. The estimation
procedure is performed in two steps. First, the observation model
parameters are estimated empirically using the observed curvature data to give $\widehat{\Sigma}_f$
and $\widehat{\mu}_f=[\widehat{\mu}_{f_1},\widehat{\mu}_{f_2},\widehat{\mu}_{f_3}]$. The type of estimation used for this first step is commonplace for data involved in Gaussian process regression \citep{Rasmussen}. Second, the parameters $\theta$ are estimated using $\widehat{\Sigma}$ and
$\widehat{\mu}_f$ and simulations from the physics-based model for $u(x)$. The full
estimation procedure is described in Algorithm \ref{alg:general}.
\begin{algorithm}
        \caption{Parameter estimation for the multi-output Gaussian Process model}\label{alg:general}
        \begin{algorithmic}[1]
            \REQUIRE Number of simulations $N_u$ from physics-based model; number of observed locations $N_f$; observed locations $X_f$; observed curvature data $\big\{z_i(X_f)\big\}_{i=1}^{M}$; regularization function
            $g(\cdot)$; physics-based model $w(\cdot,\cdot)$;

            \STATE Compute the estimates
            \begin{equation}
              \widehat{\mu}_f=\frac{1}{M}\sum^{M}_{i=1}z_i(X_{f}),
			  \label{equation:muestimates}
			  \end{equation}
			  and
			  \begin{equation}
              \widehat{\Sigma}_f =\text{diag}\left(\frac{1}{M}\sum^{M}_{i=1}\left[z_i(X_f) - \widehat{\mu}_f \right]^T \left[ z_i(X_f) - \widehat{\mu}_f\right]\right).
            \end{equation}

            \STATE Define $X_u$ to be $N_u$ evenly spaced
coordinates on $\mathcal{D}$.    
            \STATE By defining $\lambda(\theta):=\lambda=(k/4EI)^{1/4}$ for $k,EI \in \theta$, set $Y_\theta=[w(X_{u},\lambda(\theta)), \widehat{\mu}_f]
            \in\mathbb{R}^{N_u+N_f}$ and solve
\begin{equation}\widehat{\theta}=\text{arg}\max_\theta \left[L(\theta;Y_{\theta})+\log g(\theta)\right],
\label{equation:regularizedoptimization}
\end{equation}
where
\begin{equation}
L(\theta;Y_{\theta})=\frac{1}{2}\log \abs{K_{\theta}}+\frac{1}{2}Y_\theta^T K_{\theta} Y_\theta +\frac{(N_u+N_f)}{2}\log(2\pi).
\label{equation:optimizablefunction}
\end{equation}
\RETURN $\widehat{\theta} =
(\widehat{k},\widehat{EI},\widehat{\sigma^2})$ and $\widehat{\lambda}
= \left(\widehat{k} / 4 \widehat{EI} \right)^{1/4}$
\end{algorithmic}
\end{algorithm}

\label{sec:parameterestimation}

The input $g(\theta)$ is a regularization function which allows for prior belief about $\theta$ to inform the optimization. From a Bayesian perspective, (\ref{equation:regularizedoptimization}) corresponds to maximum aposteriori estimation (MAP), where the objective function $L(\theta;Y_{\theta})$ is the logarithm of the marginal likelihood $p(Y_{\theta}|X_u,X_f,\theta)$ in (\ref{equation:likelihood}). For parameters such as the flexural rigidity of a sleeper, $EI$, and ballast stiffness, $k$, there are extensive guidelines that specify confidence intervals for them \citep{PWI} and could be used to decide $g(\theta)$. Define the following log-normal distribution function,
\begin{equation}
q(x,m,s) = \frac{1}{x\sqrt{2s^2\pi}}\exp\left\{-\frac{(\log(x)-m)^2}{2s^2}\right\},
\end{equation}
then we assume that the regularization function $g(\theta)$ takes the form
\begin{equation}
g(\theta)=q(\sigma^2, 2.5, 0.125)q(k, 5, 0.25)q(EI, 28, 0.25),
\label{equation:functionforregularization}
\end{equation}
based on order-of-magnitude estimates from \cite{PWI}. This is the form of $g(\theta)$ that we consider for the remainder of the paper.


In the case of the simple physics-based model for $u(x)$ utilised in this study, the optimization problem can be solved using various iterative methods. Simulated annealing \citep{Brooks} is used for the numerical examples presented in the remainder of the paper. Let the estimated parameter vector for the model in (\ref{equation:likelihood}) be $\widehat{\theta}=(\widehat{k},\widehat{EI},\widehat{\sigma^2})$, where $\widehat{\phi}=(\widehat{k},\widehat{EI})$ and $\widehat{\lambda}=(\widehat{k}/4\widehat{EI})^{1/4}$. Then $Y_{\widehat{\theta}}=[w(X_u,\widehat{\lambda}), \widehat{\mu}_f]$ is the concatenation of the optimized physics-based model simulation and the estimated curvature. The estimation procedure above outlines the training of the proposed multi-output Gaussian process model using the batch of noisy curvature observations $\big\{z_i(X_f)\big\}_{i=1}^{M}$. Figure \ref{figure:data_and_estimated_physics} shows the same as Figure \ref{figure:data_and_physics} only with the physics-based model for the vertical deflection computed using the estimated parameters $\widehat{k}=209$ and $\widehat{EI}=6\times 10^{11}$. They were estimated using $N_u=6$ simulations from the physics-based model in (\ref{equation:optimizablefunction}). These parameter estimates lead to physics-informed curvature (via the numerically computed $\mathcal{L}_{x,\phi}w(x,\lambda)$) that is more representative of the data than that from the prior parameter beliefs in (\ref{equation:functionforregularization}).


\subsection{Generating a posterior}
\label{sec:posterior}

\begin{figure}
    \centering
    \begin{subfigure}[t]{\textwidth}
\centering
        \includegraphics[width=0.7\textwidth]{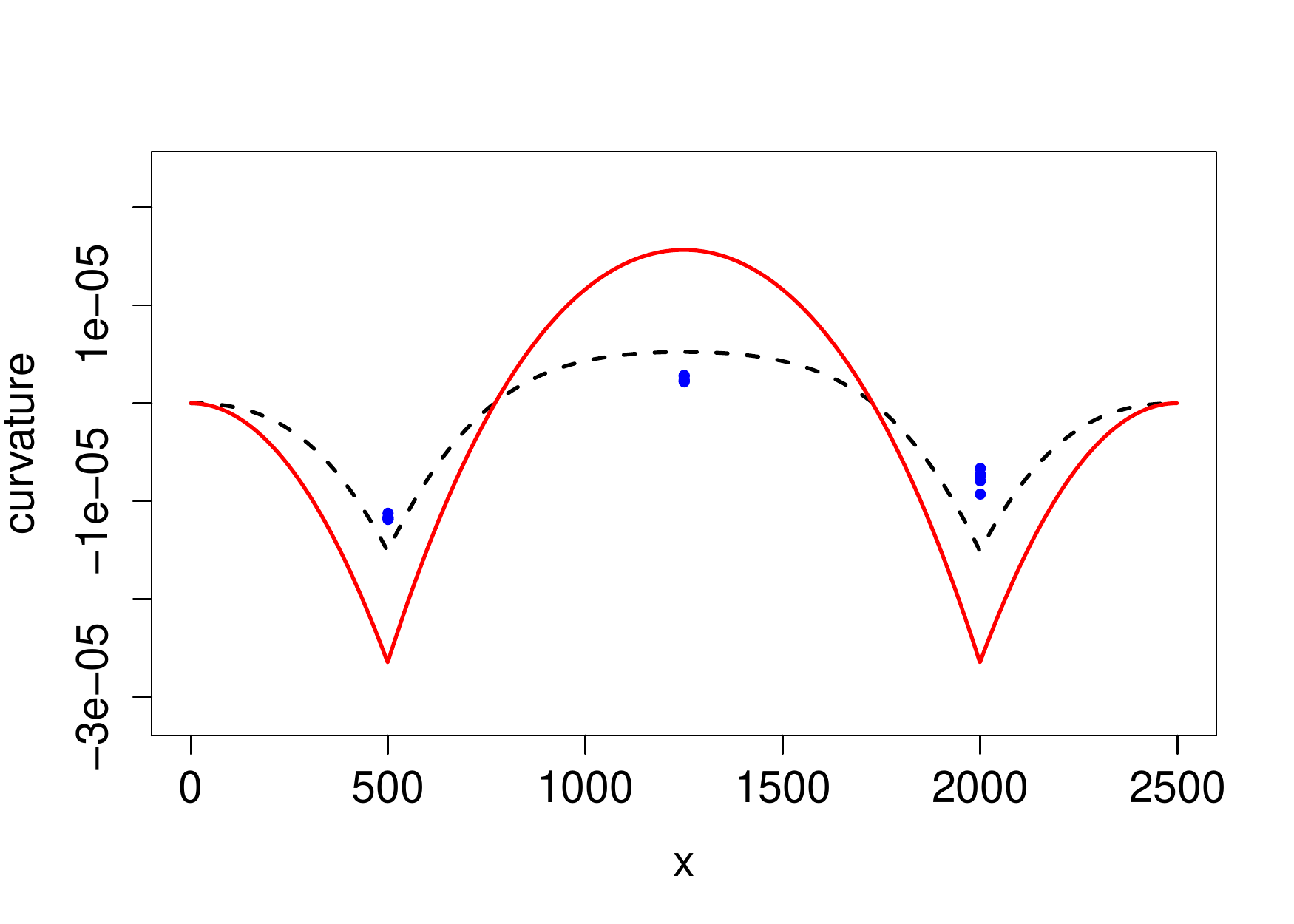}
        \caption{The simulated curvature data and model-derived curvature (red) in Figure \ref{figure:data_and_physics}, in addition to $\mathcal{L}_{\widehat{\phi},x}w(x,\widehat{\lambda})$ computed numerically from the physics-based model for the vertical deflection $y(x)$ in a dashed black line. The parameters used in the model $\widehat{\phi}$ and $\widehat{\lambda}$ are estimated through the procedure in Sec. \ref{sec:parameterestimation}.}
        \label{figure:data_and_estimated_physics}
    \end{subfigure}
	\vspace{2mm}
    ~ 
    \begin{subfigure}[t]{\textwidth}
\centering
        \includegraphics[width=0.7\textwidth]{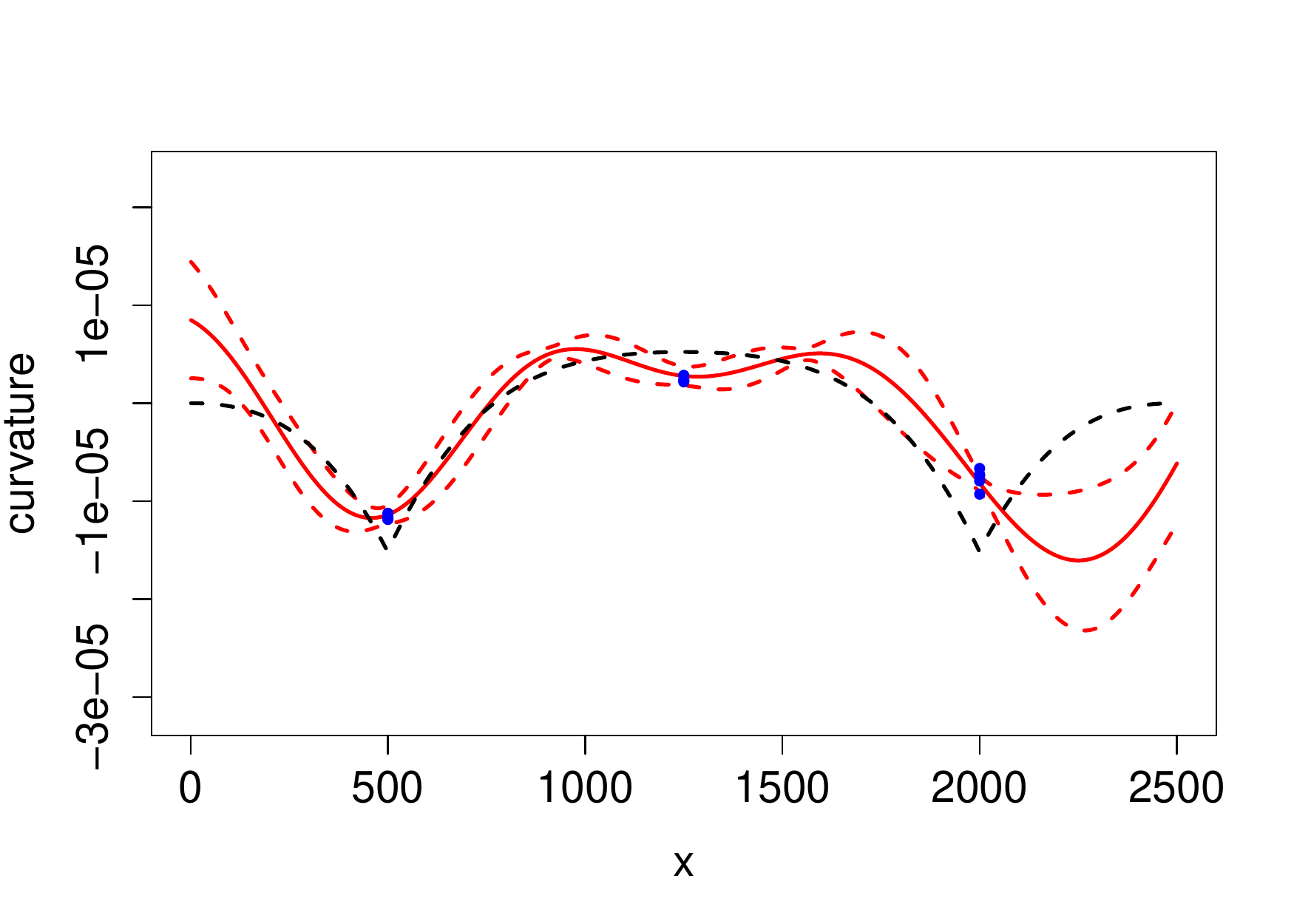}
        \caption{The same as Figure \ref{figure:data_and_estimated_physics} only with the multi-output Gaussian process posterior mean and 95\% confidence intervals as described in (\ref{equation:posteriorcurvature}).}
        \label{figure:posterior}
    \end{subfigure}
	\caption{ }
\end{figure}

Once obtaining $\widehat{\theta}$ from Algorithm \ref{alg:general}, a point-wise posterior at $x \in \mathcal{D}$ for the curvature of the sleeper given $Y_{\widehat{\theta}}$ can be computed. The posterior can be interpreted as the physics-based model guiding the inference from the sensor data in regions of the domain which are not instrumented with sensors.
Consider the joint distribution (the dependence on $X_u$, $X_f$, $\widehat{\theta}$ and $x$ is dropped for notational convenience),
$$
\begin{bmatrix}
Y_{\widehat{\theta}}\\
f(x)
\end{bmatrix} \sim \mathcal{N}\left(
\begin{bmatrix}
0\\
0
\end{bmatrix},
\begin{bmatrix}
K_{\widehat{\theta}} & k_{u,f;\widehat{\theta}}(X_u,x)\\
k_{f,u;\widehat{\theta}}(x,X_u)&k_{f,f;\widehat{\theta}}(x,x) 
\end{bmatrix}
\right).
$$
Then the posterior is given by \citep{Rasmussen,Robert},
\begin{equation}
f(x)|Y_{\widehat{\theta}},\widehat{\theta} \sim \mathcal{N}\left(q_{f}^{T}K_{\widehat{\theta}}^{-1}Y_{\widehat{\theta}},k_{f,f;\widehat{\theta}}(x,x)-q_{f}^{T}K_{\widehat{\theta}}^{-1}q_{f} \right),
\label{equation:posteriorcurvature}
\end{equation}
where
$$
q_{f}^{T}=\left(k_{f,u;\widehat{\theta}}(x,X_{u}), k_{f,f;\widehat{\theta}}(x,X_{f})\right).
$$
This posterior can be used to predict curvature (via sampling from this normal distribution), detecting system changes and to select the number of simulated data points for $u(x)$ (see next two sections). In a similar way, the posterior for the vertical deflection of the sleeper, $u(x)|Y_{\widehat{\theta}},\widehat{\theta}$, can also be obtained due to the incorporation of the physics-based model, and therefore the observed curvature data can be used to infer a physical quantity that it cannot directly measure. Notice that the posterior variance interestingly does not depend on the simulated values from the physics-based model, but only on the simulation coordinates $X_u$. As aforementioned, one could manipulate this observation by selecting the position of the simulation coordinates such that the posterior variance is minimized for future prediction; this is known as active learning \citep{Seo}. Figure \ref{figure:posterior} shows the same as Figure \ref{figure:data_and_estimated_physics} in addition to the posterior mean and 95\% confidence intervals in (\ref{equation:posteriorcurvature}).

\subsection{Tuning the quantity of simulation from $u(x)$}
\label{sec:adaptivenu}

\begin{figure}[ht!]
\centering
\includegraphics[width=100mm]{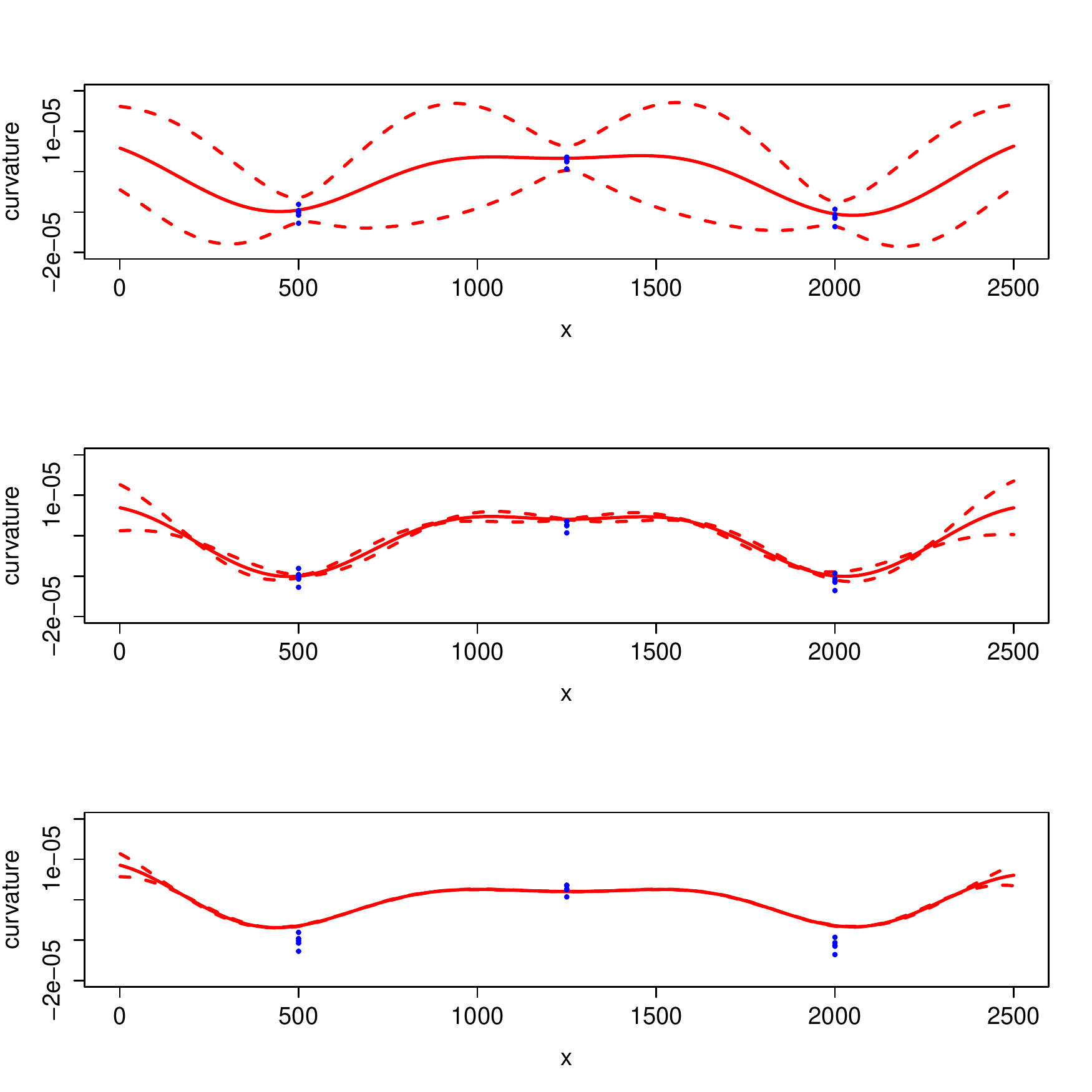}
\caption{The posteriors of curvature $p(f(x)|Y_{\widehat{\theta}},\widehat{\theta})$ corresponding to the Gaussian process models trained around $N_u=2$ (top), $N_u=9$ (middle) and $N_u=14$ (bottom) simulation coordinates from $u(x)$. The points show the simulated samples of curvature data at each point in $X_f$, and the dashed lines show the 95\% confidence interval for the posterior.}
\label{figure:goldilocks_principle}
\end{figure}

This section investigates how the similarity between the observed data and physics-based model can be assessed using the point-wise posterior for sleeper curvature. This leads to a principled method of selecting $N_{u}$, the number of simulated data-points from the analytical model for $u(x)$, that most improves the predictive performance of the posterior. In the case of an over-simplified physics-based model, we propose using more observed curvature data than the simulations from the crude physics-based model. We now demonstrate the effect that $N_u$, the number of simulations from $u(x)$, has on the posterior $p(f(x)|Y_{\widehat{\theta}},\widehat{\theta})$; this is done via simulated curvature data. Consider the simulated curvature data $\big\{z_i(X_f)\big\}_{i=1}^{5}$ sampled from
$$
z(X_f) = \begin{bmatrix}
        -1\times10^{-5}\\
    2.45\times 10^{-6}
    \\-1\times10^{-5}
  \end{bmatrix} 
+ W,\quad W \sim \mathcal{N}\left(
 0,1\times 10^{-6} I_3 \right).
$$
We take the forcing as $p=125000$N. The parameters of
the multi-output Gaussian process in (\ref{equation:likelihood}) are
estimated using $\big\{z_i(X_f)\big\}_{i=1}^{5}$ and Algorithm \ref{alg:general}, along with the number of simulated data-points set to $N_u=2$, $9$ and $14$. 
The pointwise curvature posteriors over a range of $x \in \mathcal{D}$ are then given for each model by \eqref{equation:posteriorcurvature}, and the corresponding means and 95\% confidence intervals are shown in Figure \ref{figure:goldilocks_principle}. The blue points show the simulated curvature samples at each point in $X_f$. For larger values of $N_u$ the posterior is more akin to the physics-based model than the observed data.
On the other hand, in the extreme case of $N_u=0$, the proposed method
becomes the same as standard Gaussian process regression (kriging) on
the observed curvature data; the posterior is based solely on the
observed data. As $N_u$ increases, the parameter estimates from the
associated models change accordingly, for example resulting in the
posterior diverting away from the data when $N_u=14$. In this case,
the posterior is not representative of the observed data since it
incorporates too many simulations from the physics-based model. To
obtain a posterior more representative of the observed data, fewer
simulations from the physics-based model are required, but enough that
the posterior has less variance than in the case of $N_u=2$. This
indicates that a trade-off exists between the number of simulations from
$u(x)$, $N_u$, and the predictive performance of the posterior with respect to the observed data.


The mean squared error is now employed as a metric to evaluate the data/physics information trade-off highlighted here, and to tune the number of simulations from $u(x)$. The objective here is to find the value of $N_u$ that maximises the predictive performance of the posterior in (\ref{equation:posteriorcurvature}), for a prescribed $N_u$, with respect to a test data-point. Here we suggest to use $\widehat{\mu}_{f_3}$, derived from the noisy data at $X_{f_3}$, as the test data-point. Equally this could be exchanged with either $\widehat{\mu}_{f_2}$ or $\widehat{\mu}_{f_3}$. Let $Z_{N_u}\sim f(X_{f_{3}})|Y_{N_u},\widehat{\theta}$ using (\ref{equation:posteriorcurvature}), where $Y_{N_u}=[w(X_u,\widehat{\lambda}),\widehat{\mu}_{f_1},\widehat{\mu}_{f_2}]$, and $X_u$ are $N_u$ evenly spaced coordinates along the domain. Here, $\widehat{\mu}_{f_1}$, $\widehat{\mu}_{f_2}$, $\widehat{\lambda}$ and $\widehat{\theta}$ are estimated by using Algorithm \ref{alg:general} with the inputs $N_f:=2$ and $X_f:=[X_{f_1},X_{f_2}]$ (therefore excluding the noisy curvature observations at the last sensor location). Then the mean squared error of $Z_{N_u}$ with respect to the test data-point $\widehat{\mu}_{f_3}$, $\mathbb{E}\left[(Z_{N_u}-\widehat{\mu}_{f_3})^2\right]$, is to be minimized,
\begin{equation}
\text{arg}\min_{N_u} \left\{\mathbb{E}\left[(Z_{N_u}-\widehat{\mu}_{f_3})^2\right]\right\}=\text{arg}\min_{N_u}\left\{\left(\mathbb{E}[Z_{N_u}]-\widehat{\mu}_{f_3}\right)^2+\mathbb{V}(Z_{N_u})\right\}.
\label{equation:wasserstein}
\end{equation}
The mean squared error is the sum of the squared bias of $Z_{N_u}$, with respect to $\widehat{\mu}_{f_3}$, and the variance of $Z_{N_u}$; the hope here is that the posterior $Z_{N_u}$ corresponding to the minimum in (\ref{equation:wasserstein}) will allow the physics-based model to guide it in the region where data is missing, but not be overconfident.
In practice, an interval search can be implemented between two end-points $N_u^{\min}$ and $N_u^{\max}$ to find the minimum in (\ref{equation:wasserstein}), for suitable values of $N_u^{\min}$ and $N_u^{\max}$. A summary of how this procedure interacts with the modelling framework set out in Sec. \ref{sec:parameterestimation} and \ref{sec:posterior} is
given here:
\begin{itemize}
\item[(1)] For $N_u=N_u^{\min},N_u^{\min}+1,\ldots,N_u^{\max}$, implement Algorithm \ref{alg:general} with $N_f:=2$ and $X_f:=[X_{f_1},X_{f_2}]$ as inputs to step 1 to estimate $\widehat{\theta}_{N_u}$.
\item[(2)] For each $\widehat{\theta}_{N_u}$ compute the posterior in (\ref{equation:posteriorcurvature}) and record the mean squared error $\mathbb{E}\left[(Z_{N_u}-\widehat{\mu}_{f_3})^2\right]$ where $Z_{N_u}\sim f(X_{f_3})|Y_{N_u},\widehat{\theta}_{N_u}$.
\item[(3)] Choose the value of $N_u$ corresponding to the lowest mean squared error.
\item[(4)] Implement Algorithm \ref{alg:general} with this value of $N_u$, and by using $N_f:=3$ and $X_f:=[X_{f_1},X_{f_2},X_{f_3}]$ as inputs to step 1, to obtain the posterior in (\ref{equation:posteriorcurvature}).
\end{itemize}

\begin{figure}[ht!]
\centering
\includegraphics[width=100mm]{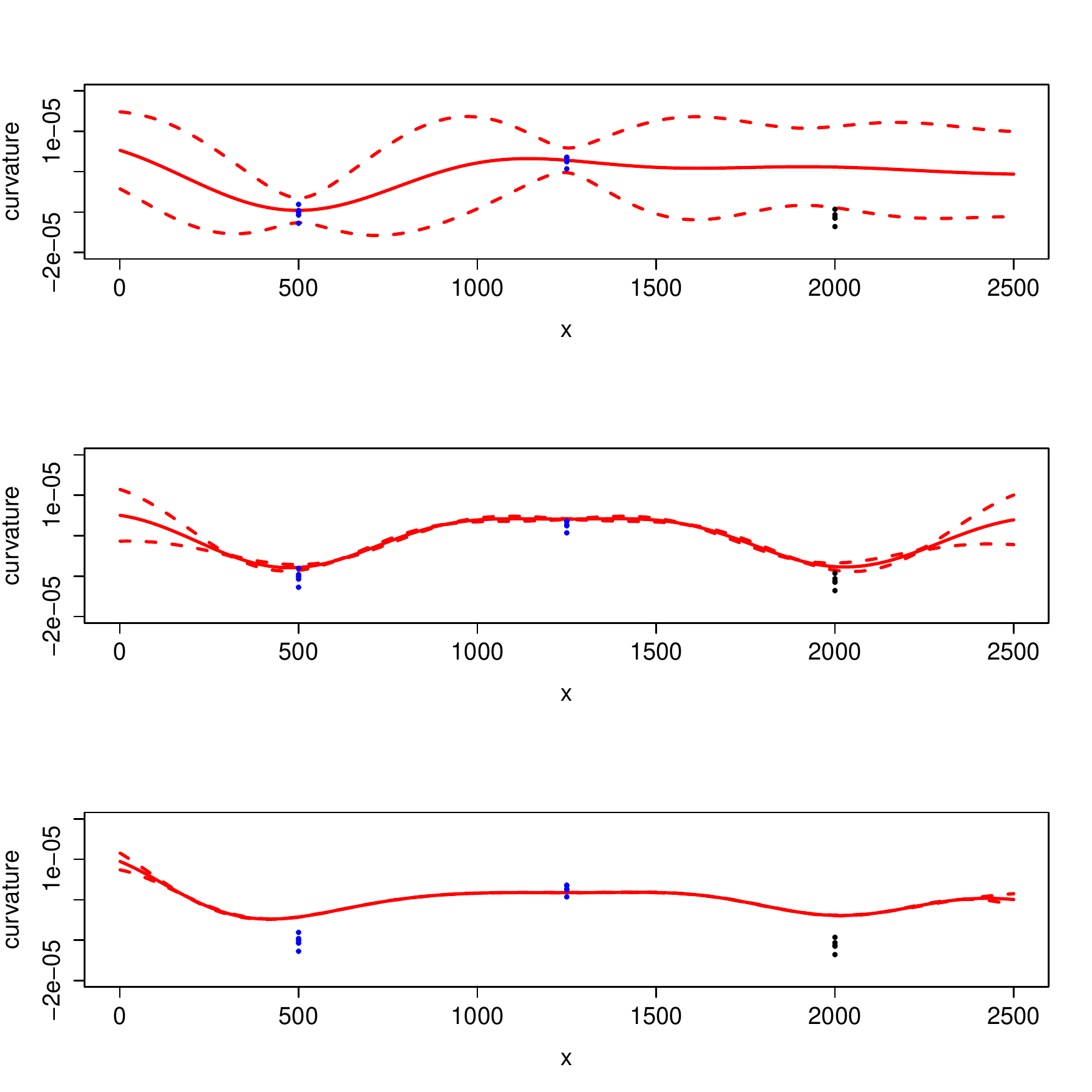}
\caption{The posteriors $p(f(x)|Y_{N_u},\widehat{\theta})$ corresponding to the same simulated curvature data as used in Figure \ref{figure:goldilocks_principle}, for $N_u=2$ (top), $N_u=9$ (middle) and $N_u=14$ (bottom).}
\label{figure:goldilocks_principle_2}
\end{figure}

\begin{figure}[ht!]
\centering
\includegraphics[width=100mm]{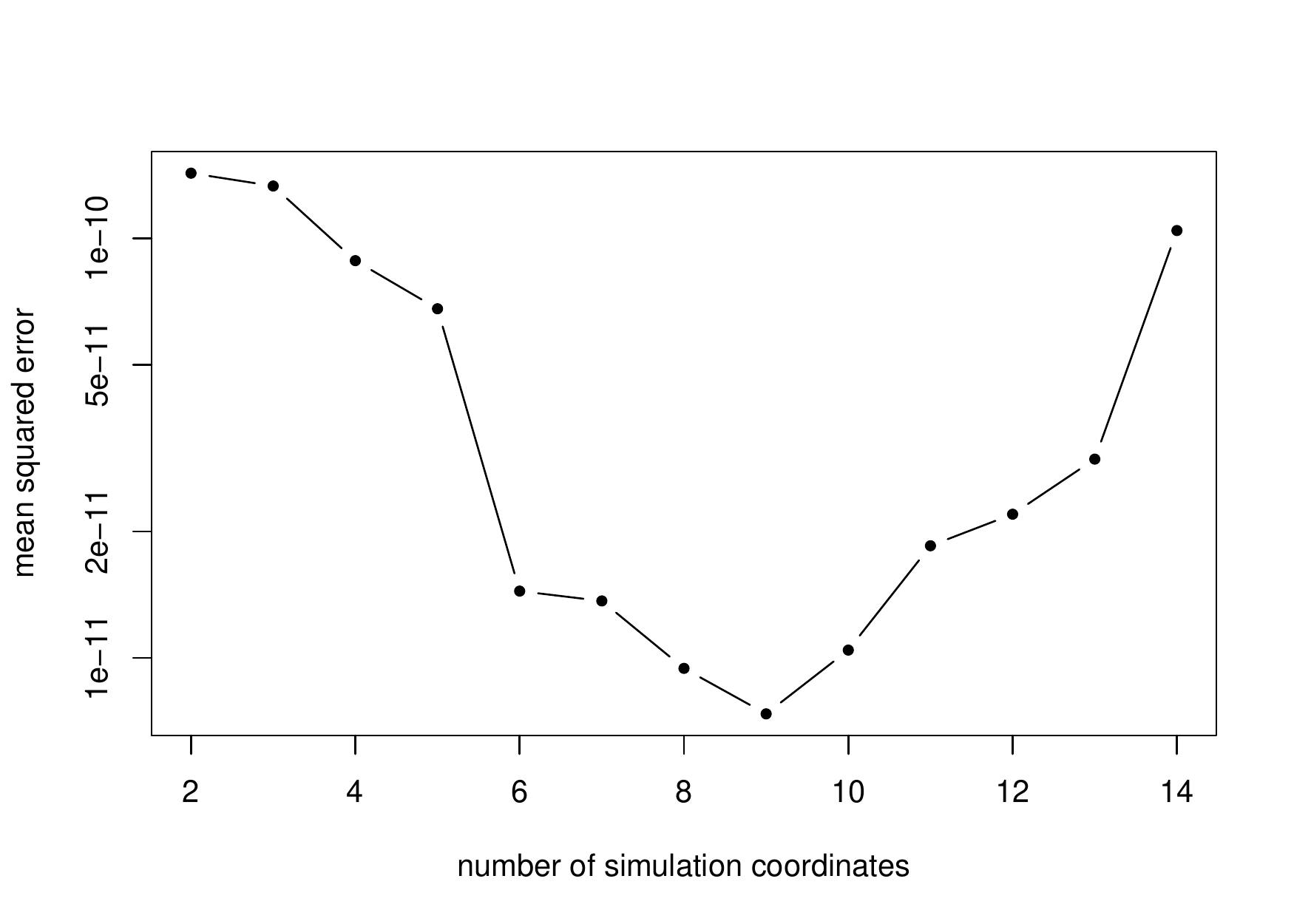}
\caption{The mean squared error of the posteriors $p(f(X_{f_3})|Y_{N_u},\widehat{\theta})$ with respect to $\widehat{\mu}_{f_3}$, for the range of $N_u$ values between 2 and 14.}
\label{figure:mean_squared_error}
\end{figure}

Figure \ref{figure:goldilocks_principle_2} shows the posteriors $p(f(x)|Y_{N_u},\widehat{\theta})$ for the same simulated curvature data as used in Figure \ref{figure:goldilocks_principle}, with $N_u=2$, $N_u=9$ and $N_u=14$ again. The simulated curvature samples at the last coordinate in $X_f$, that are excluded from the training of the models, are shown here. One can see the effect of the additional simulation coordinates, for the posteriors corresponding to $N_u=9$ and $N_u=14$, in imparting structure from the physics-based model. Figure \ref{figure:mean_squared_error} shows the mean squared error for the posteriors corresponding to the range of $N_u$ values between 2 and 14 (again computed using the same simulated curvature data as above); this quantifies the predictive performance of the posteriors shown in Figure \ref{figure:goldilocks_principle}. By using the mean squared error to tune $N_u$ in the way described above, the value of $N_u$ could be used as a proxy to evaluate how well the data fits the physics-based model.

Once $N_u$ has been prescribed, the choice of the coordinates $X_u$ has to be decided. For the remainder of the paper they are chosen to be distributed uniformly along the entire domain $\mathcal{D}$. For a more advanced scheme, one could use active learning techniques \cite{Seo}, which prescribes simulation coordinates that minimise the variance of aposteriori estimates. In addition to this, knowledge of where the curvature data are available, $X_f$, could lead to a wiser decision on the choice of simulation coordinates $X_u$. For example if all data were known to come from only a single region of the domain, simulations should be taken at points outside this region as to guide the inference from the data in these areas as much as possible.

\subsection{Detection of system changes}
\label{sec:damagedetection}

This section will discuss the use of the methodology presented in the previous sections for detecting changes in the modelled system and the parameters within the physics-based model. In terms of the application of this paper, an underlying change in the properties of the sleeper (and it's foundation) could signify potential damage. Therefore the methods discussed in this section can be used for structural damage detection.
The aim of detecting system changes is to signal when the underlying physics (e.g. model parameters) has changed. To achieve this we use batches (length $B$) of curvature data $\mathcal{Z}_i=\big\{z_j(X_{f})\big\}_{j = iB+1}^{i(B+1)}$, for $i=0,1,2,3,\ldots$. Then $\widehat{\mu}_f^{(i)}$ are the sample means, in (\ref{equation:muestimates}), of the batches. This can be implemented alongside the proposed methodology in this paper via two general approaches:
\begin{itemize}
\item[]\textbf{Change in estimated parameters:} Fit separate Gaussian process models for the curvature data $\mathcal{Z}_i$, for $i=0,1,2,3,\ldots$ using Algorithm \ref{alg:general}, and find estimates for $\theta$ denoted by $\widehat{\theta}^{(i)}$. Then statistically compare $\widehat{\theta}^{(i)}$ with $\widehat{\theta}$.

\item[]\textbf{Change in log marginal likelihood:}
Let $Y^{(i)}=[w(X_u,\widehat{\lambda}),\widehat{\mu}^{(i)}_f]$ and find the log marginal likelihood
$\log p(Y^{(i)}|X_u,X_f,\widehat{\theta}):=L(\widehat{\theta};Y^{(i)})$.
Statistically compare this to $\log p(Y_{\widehat{\theta}}|X_u,X_f,\widehat{\theta})$.
\end{itemize}
The former technique is commonly utilised in model updating, for example by using the relative error of $\widehat{\theta}^{(i)}$ with respect to $\widehat{\theta}$ as a metric to detect significant parameter changes over time \citep{Schommer}.
On this note, many popular unsupervised change-point detection algorithms can be used alongside the two approaches mentioned to identify a sudden change in parameter estimates/marginal likelihood, such as those presented in \cite{Lau, Gregory}. There are many aspects of these methods to consider, such as the uncertainty in the hypothesis test statistics and the choice of any hard-thresholds. For the results presented in the next section, we will use the metrics discussed here to assess changes in the modelled system over time.

\section{Results}
\label{sec:results}

The following sections contain example implementations and results of the proposed methodology, for simulated and experimental data under laboratory conditions. These examples both consist of system change detection problems discussed in the previous section.

\subsection{Simulated curvature data}
\label{sec:simulatedresults}


This section will demonstrate the effectiveness of the proposed methodology for detecting changes in the modelled system. A data set of 100 data-points are simulated, the first 50 curvature data-points are generated using the values $EI_0$ and $k_0$ for the flexural rigidity and the ballast stiffness respectively, and the remaining 50 are generated using the values $EI_1$ and $k_1$. Further we set the force value $p=125000$N. The simulated curvature data, $z_j(X_f)$, for $j=1,\ldots,100$, is generated via perturbations to a finite-difference approximation to (\ref{equation:curvature}), namely for all $x \in X_f$,
\begin{equation}
z_j(x)=
\begin{cases}
\begin{split}
\frac{p\lambda_0}{k_0}\big\{&w((x+1),\lambda_0)-2w(x,\lambda_0)+...\\
\quad & w((x-1),\lambda_0)\big\}
+ \xi,\end{split} &\text{for } j= 1,\ldots,50 \\
\begin{split}
\frac{p\lambda_1}{k_1}\big\{&w((x+1),\lambda_1)-2w(x,\lambda_1)+...\\
\quad & w((x-1),\lambda_1)\big\}
+ \xi,\end{split} &\text{for } j= 51,\ldots,100,
\end{cases}
\label{equation:simulateddataresults}
\end{equation}
where $\xi \sim \mathcal{N}(0, 5\times10^{-7})$ and $\lambda_0=(k_0/(4EI_0))^{\frac{1}{4}}$ and $\lambda_1=(k_1/(4EI_1))^{\frac{1}{4}}$.
The proposed multi-output Gaussian process model is trained using the
first batch (length $B=5$) of data, $\mathcal{Z}_{0}$, in Algorithm \ref{alg:general}. In order to quantify the uncertainty in detecting simulated parameter changes, given the uncertainty in the data-points, the log marginal likelihood and parameter estimates in the following sections will be computed over 20 independently sampled data sets.

\subsubsection{Changing the flexural rigidity}
\label{sec:flexuralrigidity}

In the first case we set $k_0=k_1=450$, $EI_0=8 \times 10^{11}$ and $EI_1=6 \times 10^{11}$; these values are of the same order of magnitude to those prior beliefs (see \ref{equation:functionforregularization}) for the sleeper system considered in this paper. Thus the flexural rigidity is reduced at the midpoint of the simulation and the ballast stiffness is kept fixed. A simulated reduction, as supposed to an increase, in the flexural rigidity is used to represent the cracking of concrete. The tuning procedure outlined in Sec. \ref{sec:adaptivenu} is used to specify the value of $N_u$ used for the training of the proposed model and this results in $N_{u}=9$. Using the training data $\mathcal{Z}_0$ and Algorithm \ref{alg:general}, the estimates $\widehat{\theta}=[\widehat{k},\widehat{EI},\widehat{\sigma^{2}}]=[286.3, 5.0 \times 10^{11}, 13.6]$ are obtained. Due to the given variance in the simulated data-points from (\ref{equation:simulateddataresults}), it is not expected that the estimates will be exactly the values prescribed, $k_0$ and $EI_0$.

\begin{figure}[ht!]
\centering
\includegraphics[width=100mm]{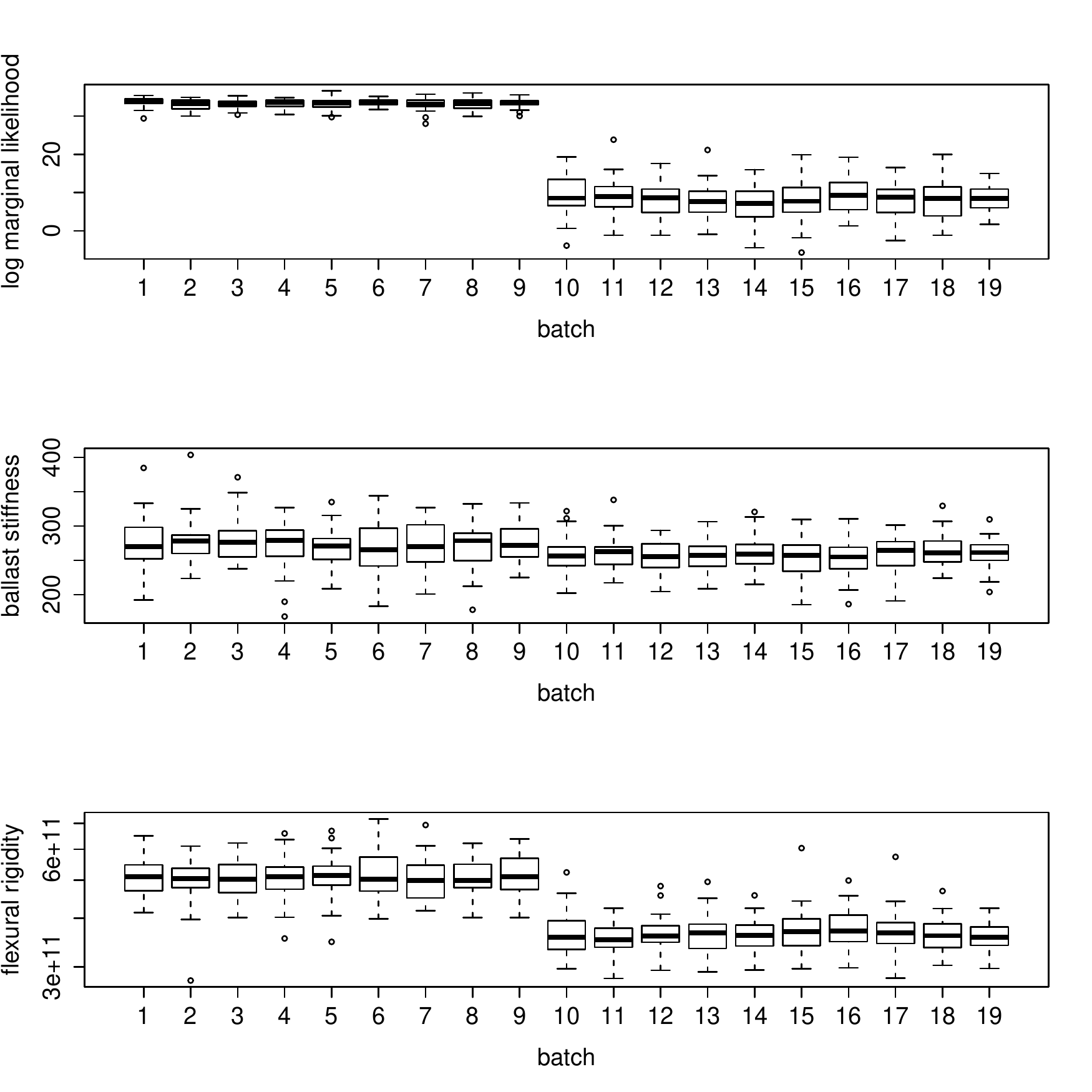}
\caption{The box-plots (over 20 independently sampled data sets) of the log marginal likelihood $\log p(Y^{(i)}|\widehat{\theta})$ and estimates $\widehat{k}^{(i)}$ and $\widehat{EI}^{(i)}$ for each batch $\mathcal{Z}_i$ ($i=1,...,19$) of simulated curvature data points. As the true change in $EI$ occurs after the 50'th data point, the likelihood decreases.}
\label{figure:curvature_likelihood_sim_data}
\end{figure}

Next, the log marginal likelihood $\log p(Y^{(i)}|\widehat{\theta})$ is computed, recalling that $Y^{(i)}=[w(X_u,\widehat{\lambda}),\widehat{\mu}_f^{(i)}]$ from Sec. \ref{sec:damagedetection}, where $\widehat{\mu}^{(i)}_f$ is the sample mean in (\ref{equation:muestimates}) of the batch $\mathcal{Z}_i$, for $i=1,...,19$. Figure \ref{figure:curvature_likelihood_sim_data} shows these log marginal likelihoods over all batches of data (and all 20 sampled data sets). Also shown in this figure are values of $\widehat{k}^{(i)}$ and $\widehat{EI}^{(i)}$ (over all data sets) obtained by fitting separate Gaussian process models on each batch of simulated data points $\mathcal{Z}_i$, for $i=1,\ldots,19$, using Algorithm \ref{alg:general}. Interestingly here, the results correctly detect that $EI$ is reduced after the true change occurs at the 50'th data point and that $k$ stays approximately constant throughout. Also note the clear decrease in likelihood after the true change in $EI$ occurs.

\subsubsection{Changing the ballast stiffness}
\label{sec:ballaststiffness}

\begin{figure}[ht!]
\centering
\includegraphics[width=100mm]{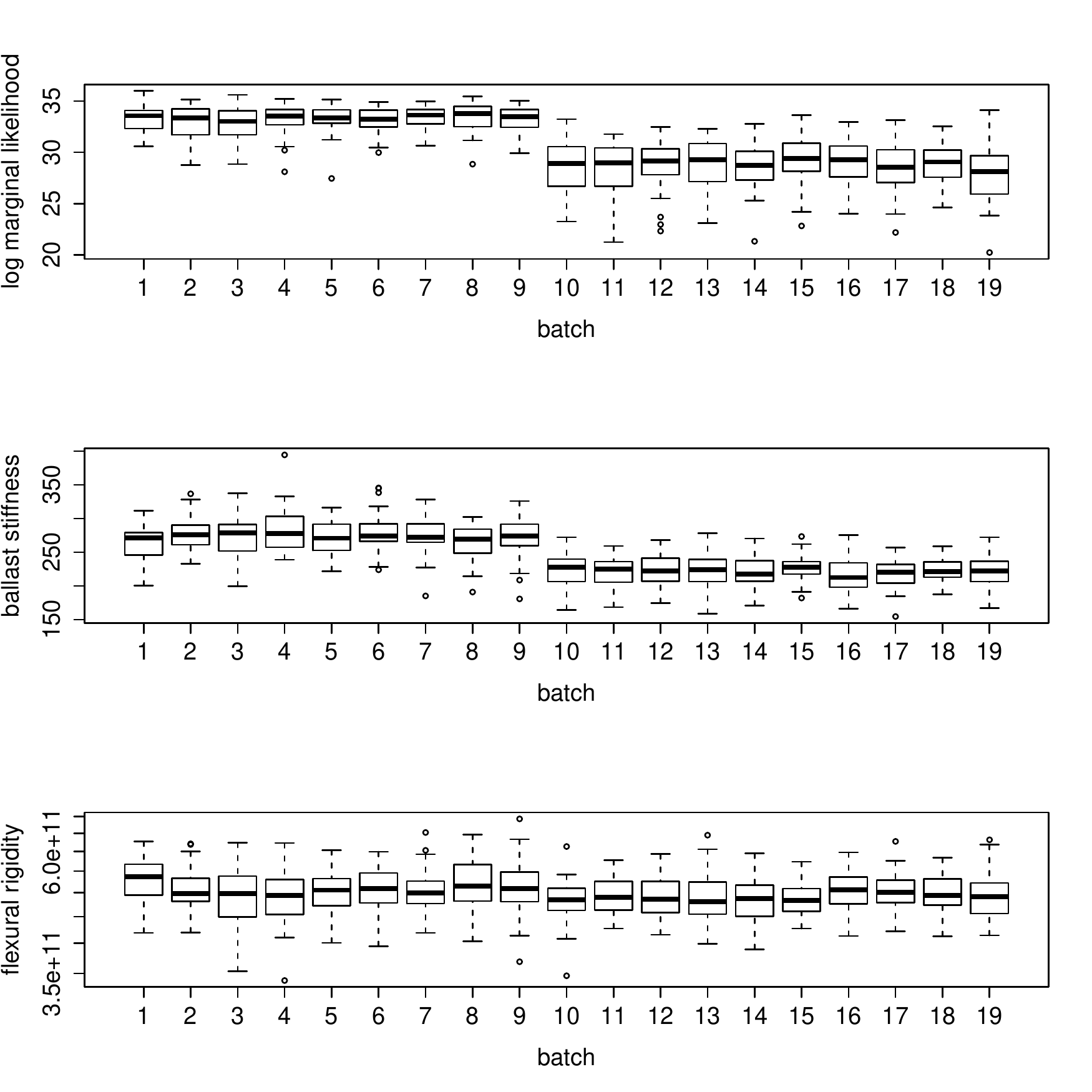}
\caption{Same as Figure \ref{figure:curvature_likelihood_sim_data} only for a change in $k$ for the second set of 50 simulated data points.}
\label{figure:curvature_likelihood_k_change_sim_data}
\end{figure}

In the second case we set $EI_0=EI_1=8 \times 10^{11}$, $k_0=450$ and $k_1=300$. Thus the ballast stiffness now changes at the midpoint of the simulation and the flexural rigidity is kept fixed. As done in the previous case with varying flexural rigidity, the log marginal likelihoods $\log p(Y^{(i)}|\widehat{\theta})$ in addition to the estimates $\widehat{k}^{(i)}$ and $\widehat{EI}^{(i)}$ are obtained for each batch of simulated curvature data points (over 20 independently sampled data sets) $\mathcal{Z}_i$, for $i=1,\ldots,19$. These quantities are all shown in Figure \ref{figure:curvature_likelihood_k_change_sim_data}. Once again the likelihoods and the values of $\widehat{k}^{(i)}$ significantly decrease after the true change in $k$ occurs. In addition to this behaviour, the new estimates $\widehat{EI}^{(i)}$ also stay approximately constant as expected.

\subsubsection{Receiver operating characteristic analysis}

\begin{figure}[ht!]
\centering
\begin{minipage}{\textwidth}
  \centering
  \vspace{8mm}
  \includegraphics[width=85mm]{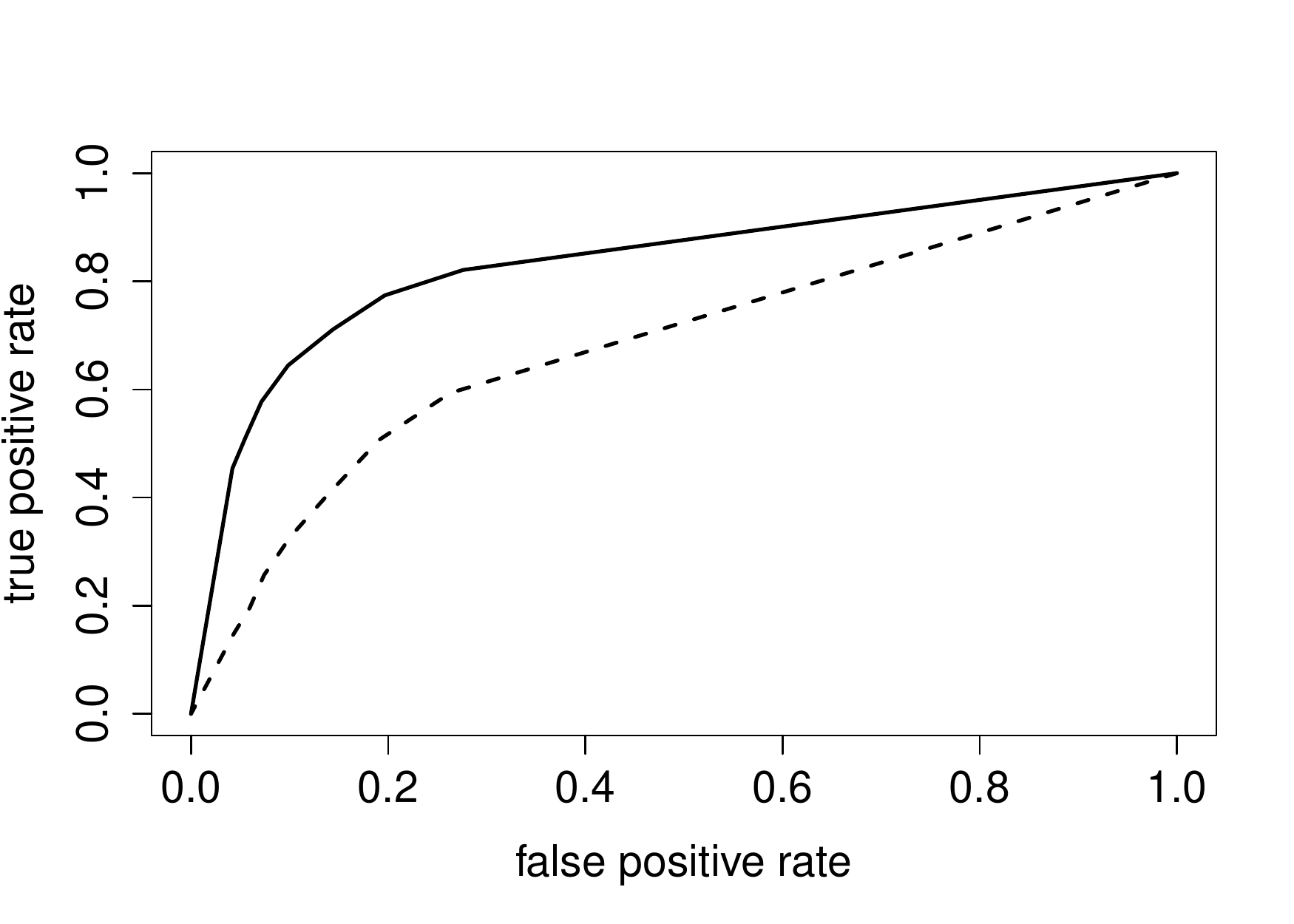}
\captionof{figure}{ROC curves showing the true positive detection rate (TPR) and false positive detection rate (FPR) in (\ref{equation:rates}). They are computed for the parameter change combinations $(EI_1,k_1)$ of $(7.8 \times 10^{11}, 435)$, in red, and $(7.8 \times 10^{11}, 420)$, in black, that occur in the second set of 50 simulated data points.}
\label{figure:roc_curves}
\end{minipage}%
\vspace{2mm}
\begin{minipage}{\textwidth}
  \centering
\includegraphics[width=95mm]{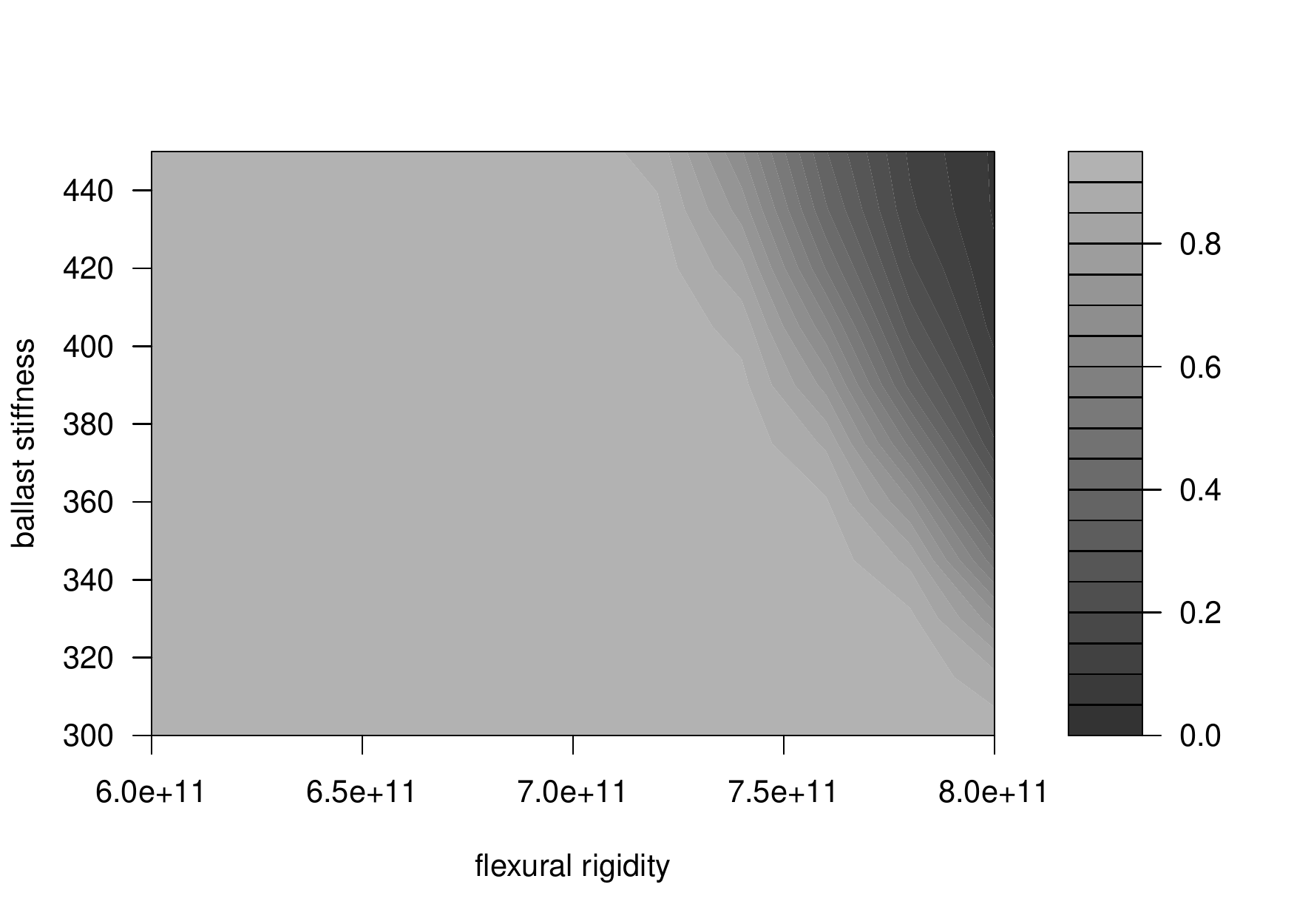}
\captionof{figure}{The area under the ROC curves showing the true positive detection rate (TPR) and false positive detection rate (FPR) in (\ref{equation:rates}). They are computed for all of the parameter change combinations that occur in the second set of 50 simulated data points.}
\label{figure:auc}
\end{minipage}
\end{figure}

\begin{figure}[ht!]
\centering
\includegraphics[width=140mm]{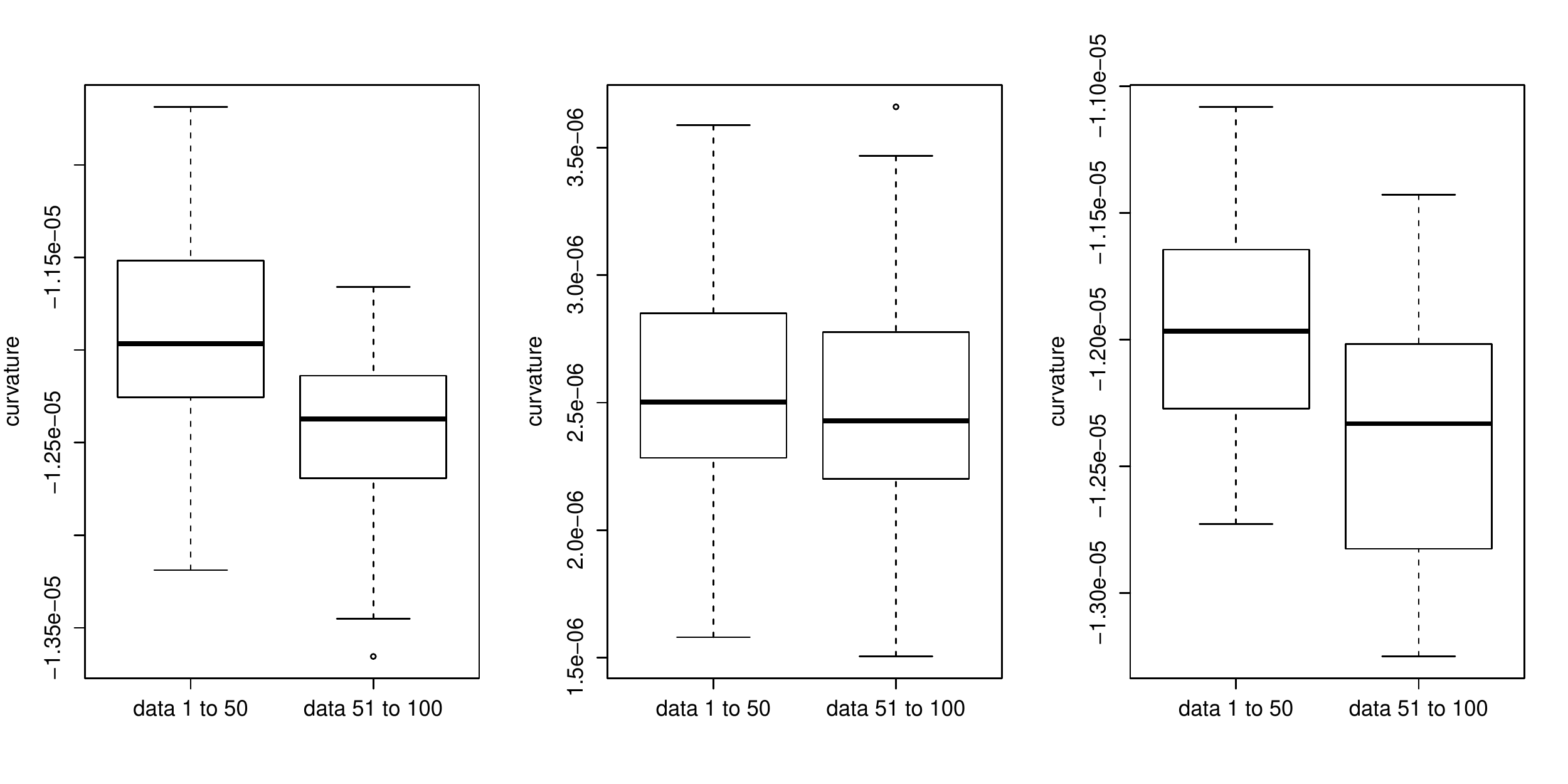}
\caption{The boxplots of the simulated curvature data in (\ref{equation:simulateddataresults}) for both $j \in [1,\ldots,50]$ and $j \in [51,100]$ and the parameters $EI_0=8 \times 10^{11}$, $k_0=450$, $EI_{1}=0.9 EI_0$ and $k_1=0.9k_0$. These are shown for $X_{f_1}$ (left), $X_{f_2}$ (middle) and $X_{f_3}$ (right).}
\label{figure:sim_curvature_data}
\end{figure}

A receiver operating characteristic (ROC) analysis \citep{Fawcett}
will now be carried out to demonstrate the sensitivity of the log marginal likelihood shown in both Figures \ref{figure:curvature_likelihood_sim_data} and \ref{figure:curvature_likelihood_k_change_sim_data} to any changes in the parameters $EI$ and
$k$ during the second set of 50 data points. This analysis evaluates
how large the change in the parameters is required to be in order to
produce a significant shift in the log marginal likelihood. The parameters
are changed from $k_0=450$ and $EI_0=8 \times 10^{11}$ to $k_1$ and
$EI_1$ respectively after the 50'th simulated data point as done in
Sec. \ref{sec:simulatedresults}, and are varied as follows:
$EI_1 \in [6 \times 10^{11}, 6.2 \times 10^{11}, 6.4 \times 10^{11},
\ldots, 8 \times 10^{11}]$ and $k_1 \in [300, 315, 330, \ldots,
450]$. Therefore there are $11^2$ different parameter combinations
after the change in this analysis. We now define the signalling
  procedure, based on the log marginal likelihood $p(Y^{(i)}|\widehat{\theta})$, to flag changes in
  the model parameters. Fitting a normal distribution to the
log marginal likelihood $\log p(Y_{\widehat{\theta}}|\widehat{\theta})$ from using the first batch of data points, $\mathcal{Z}_0$, to train the
Gaussian process through Algorithm \ref{alg:general}, we define
$\mu_0$ and $\sigma_0$ to be the empirical mean and standard
deviation over 1000 independent data sets in (\ref{equation:simulateddataresults}). Then the test-statistic for the signalling procedure we
use is based on the log marginal likelihood
$\log p(Y^{(i)}|\widehat{\theta})$, for
$i=1,...,19$, and is given by
$$
p_i=\Phi\left( \log p(Y^{(i)}|\widehat{\theta}) |\mu_0, \sigma_0\right),
$$
where $\Phi(\cdot|\mu_0,\sigma_0)$ is a Gaussian distribution function with mean $\mu_0$ and standard deviation $\sigma_0$. Then using a threshold $\gamma$ to signal a change in a log marginal likelihood, define the true positive detections as an actual change in the true parameters when one is signalled; i.e. $TP(\gamma)=\lvert \big\{i \in [10, 19]; p_i < \gamma\big\} \rvert/10$. Similarly, define the false positive detections as no actual change in the true parameters when one is signalled; i.e. $FP(\gamma)=\lvert \big\{i \in [0, 9]; p_i < \gamma\big\} \rvert/10$, where $p_0=\Phi\left(\log p(Y_{\widehat{\theta}}|\widehat{\theta})|\mu_0,\sigma_0\right)$. For a fixed parameter combination the detection rates are computed over 1000 independent data sets, $TP_j(\gamma)$ and $FP_j(\gamma)$, for $j=1,\dots,1000$:
\begin{equation}
TPR(\gamma)=\frac{\sum^{1000}_{j=1}TP_j(\gamma)}{1000}, \quad FPR(\gamma)=\frac{\sum^{1000}_{j=1}FP_j(\gamma)}{1000}.
\label{equation:rates}
\end{equation}
A ROC curve is the graph of $TPR(\gamma)$ against $FPR(\gamma)$ for varying $\gamma \in [\mu_0-0.5 \sigma_0,\mu_0-0.75 \sigma_0,\mu_0-\sigma_0,\ldots,\mu_0-2 \sigma_0]$. Figure \ref{figure:roc_curves} shows these curves for the parameter change combinations $(EI_1, k_1)$ of $(7.8 \times 10^{11}, 435)$ and $(7.8 \times 10^{11}, 420)$. The area under each curve (AUC), $\sum_{\gamma} TPR(\gamma)[1-FPR(\gamma)]$, in this case represents the probability of the detecting a random instance of a change in the parameters with respect to the test statistic used. The AUC for all parameter combinations is summarized in Figure \ref{figure:auc}. This plot shows the magnitude of parameter changes that the detection utilising the log marginal likelihood is sensitive to; this includes the values of $EI_1$ and $k_0$ prescribed in Sec. \ref{sec:flexuralrigidity} and \ref{sec:ballaststiffness} respectively. The ROC analysis presented here has allowed us to evaluate the sensitivity of the log marginal likelihood from the proposed multi-output Gaussian process model to parameter changes within the physics-based model.
For example, the AUC suggests that a true decrease of 10\% in both parameters, i.e. $EI_1=0.9EI_0$ and $k_1=0.9k_0$, is 80\% likely to be detected. Figure \ref{figure:sim_curvature_data} show boxplots for the simulated curvature data, for each location in $X_f$, in (\ref{equation:simulateddataresults}) corresponding to this 10\% decrease in the parameters. Note that the 10\% change in the true parameters is difficult to detect by visual inspection, however the signalling procedure detects this change with high probability.

\subsection{Experimental curvature data}
\label{sec:experiment}

\begin{figure}[ht!]
\centering
\begin{minipage}{\textwidth}
  \centering
  \vspace{9mm}
  \includegraphics[width=105mm]{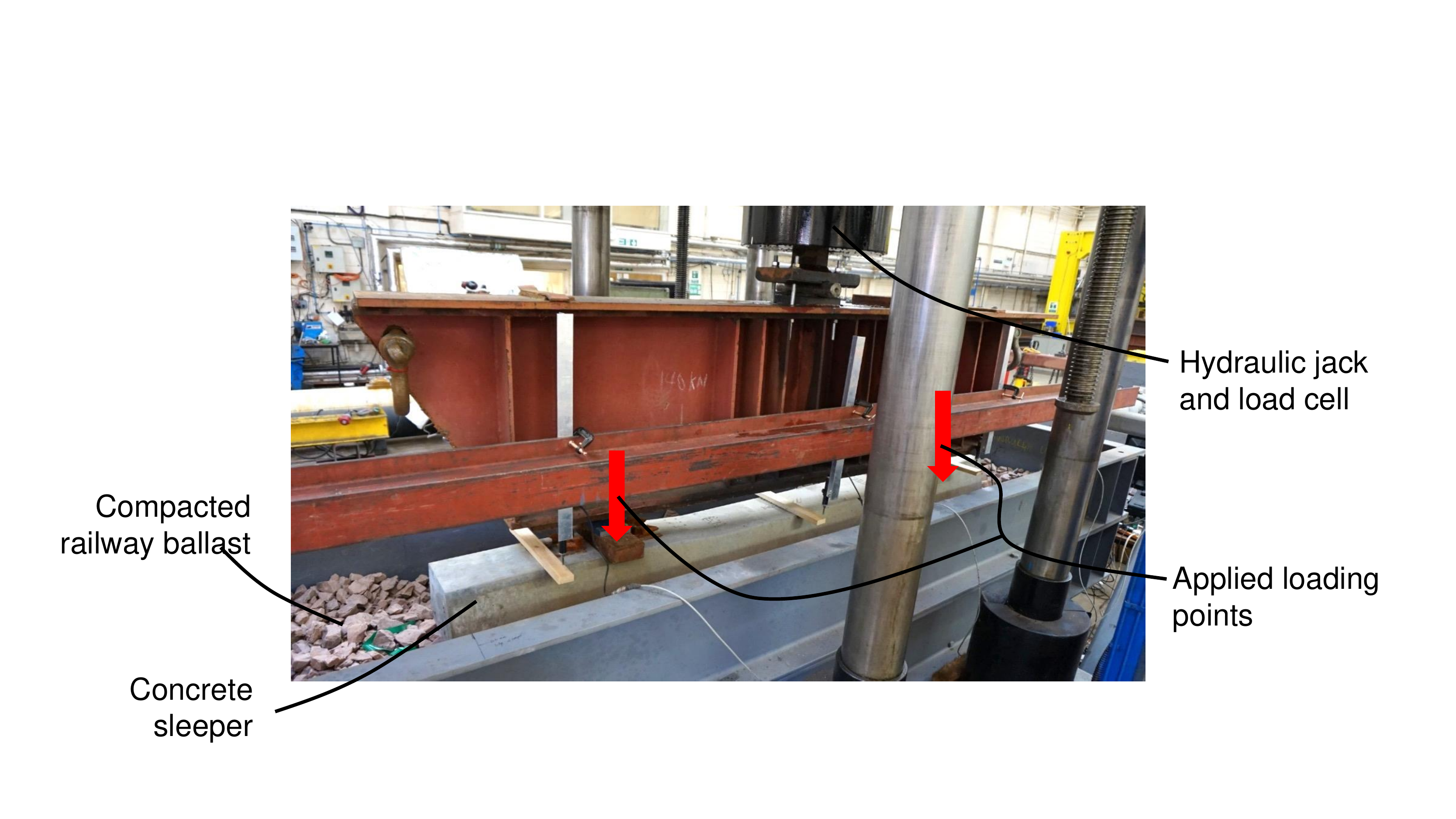}
\captionof{figure}{The experimental setup for the testing procedure in Sec. \ref{sec:experiment}. This shows the actuator/hydraulic jack applying a prescribed forcing to a spreader beam that sits above two loading points $x_1$ and $x_2$ on the sleeper beam.}
\label{figure:schematic4}
\end{minipage}%
\vspace{2mm}
\begin{minipage}{\textwidth}
  \centering
\includegraphics[width=90mm]{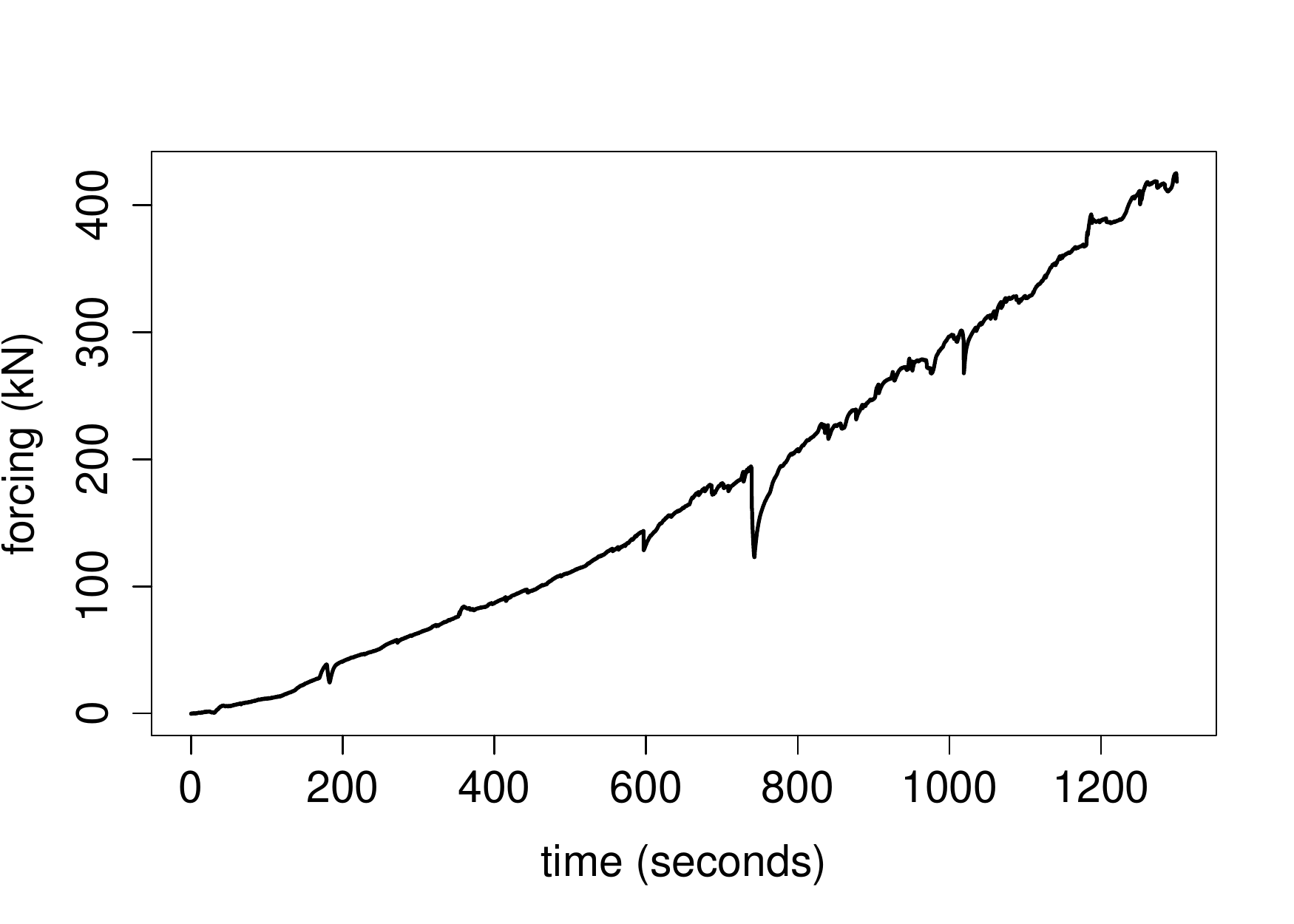}
\captionof{figure}{The actuator forcing over the entire testing procedure. This forcing is exerted on to the loading points on the sleeper via a spreader beam.}
\label{figure:forcing}
\end{minipage}
\end{figure}

This section demonstrates the effectiveness of the proposed technique when applied to the experimental data. The laboratory experiment considered was introduced in \cite{Butler}. The experiment consisted of a single instrumented railway sleeper resting in the centre of a 3m wide and 0.4m deep ballast test-bed. This test setup was designed to replicate the conditions under which the sleeper would exhibit within an actual railway bed. The sleeper has the same specifications and sensor locations as considered throughout this paper (see Figure \ref{figure:schematic}). A 0.6m deep steel spreader beam was positioned above the sleeper that applied a vertical actuator/hydraulic jack load through two rigid loading points (at $x_1$ and $x_2$) mounted on the sleeper. The experimental setup is shown in Figure \ref{figure:schematic4}. The testing procedure involved applying a set of two equal vertical forces at a linearly increasing rate up until the point where damage (in the form of concrete cracking and/or significant ballast settlement) was visually observed. The entire test was completed in approximately 1300 seconds. The actuator forcing for the testing procedure is shown in Figure \ref{figure:forcing}. For full specifications of the experiment and testing procedure, turn to \cite{Butler}.

\begin{figure}[ht!]
\centering
\begin{minipage}{0.9\textwidth}
  \centering
  \vspace{8mm}
  \includegraphics[width=85mm]{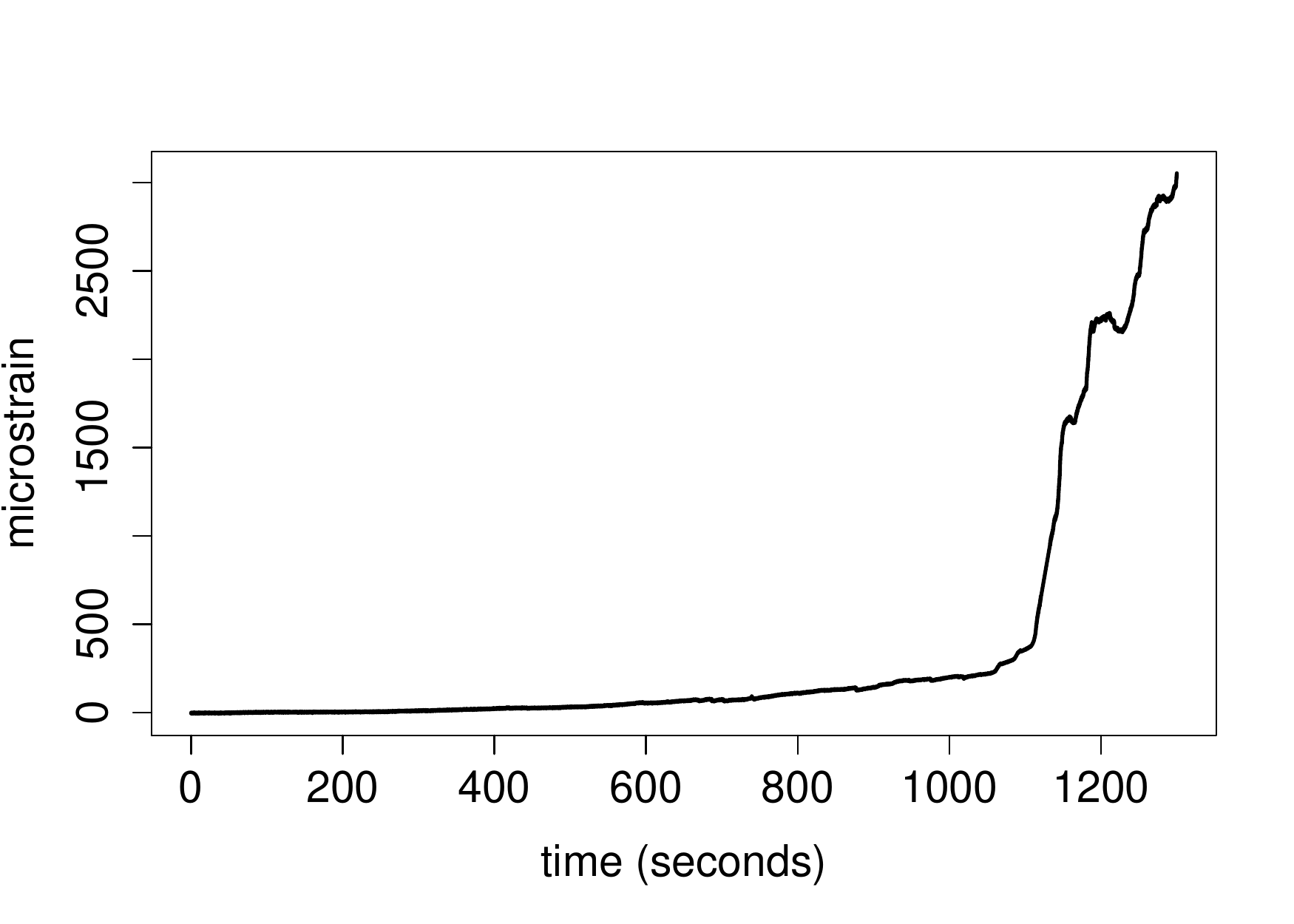}
\captionof{figure}{Strain measurements from the FBG sensor located at $x=d/2$ on the top prestressing strand, embedded in the experimental sleeper, during the testing procedure. There is a large increase in strain recorded between 1100 and 1150 seconds suggesting structural change in the sleeper.}
\label{figure:strain_sleeper_experiment}
\end{minipage}%
\vspace{2mm}
\begin{minipage}{0.9\textwidth}
  \centering
\includegraphics[width=85mm]{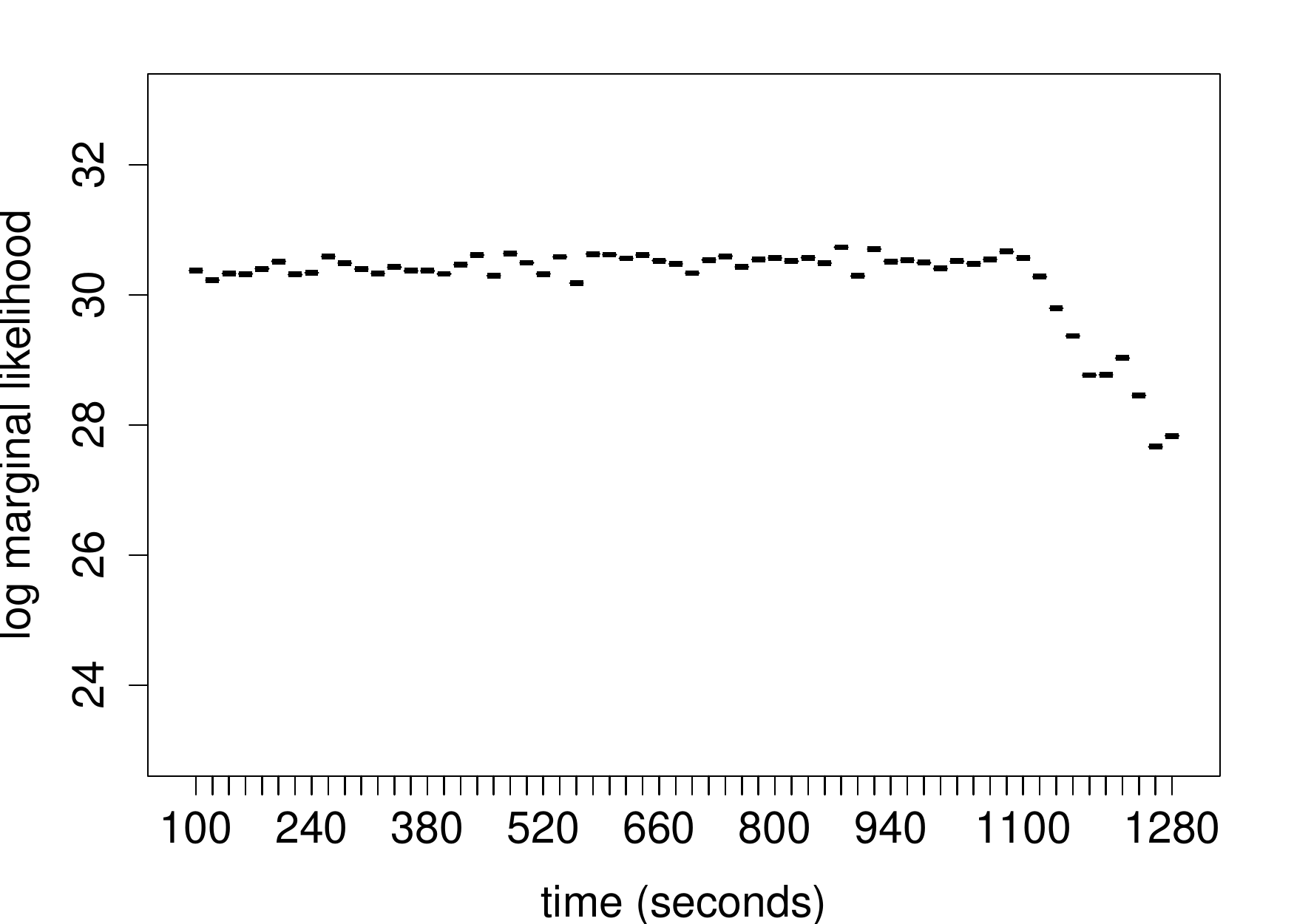}
\captionof{figure}{The log marginal likelihood $\log p(Y^{(i)}|\widehat{\theta})$ of each 1000-wide batch of curvature data points given the parameters $\widehat{\theta}$ estimated during the first 1000 data points. This likelihood sharply decreases between 1100 and 1120 seconds.}
\label{figure:log_likelihood_sleeper_experiment}
\end{minipage}
\end{figure}

The aim of this experiment is to estimate the flexural rigidity of the sleeper and the stiffness of the ballast beneath, and to obtain a posterior model for the curvature of the sleeper beam using the first 100 seconds of curvature from the instrumented sleeper. This is done using the technique proposed in this paper. From this model, the objective is then to detect at which time throughout the testing procedure a change to the system (e.g. reduction in flexural rigidity resulting from concrete cracking or reduction in ballast stiffness due to settlement/crushing) occurs. The approach taken for this detection is the second one listed in Sec. \ref{sec:damagedetection}. For this example, the actuator forcing $p$ is known (shown in Figure \ref{figure:forcing}). Figure \ref{figure:strain_sleeper_experiment} shows the strain measurements from the FBG sensor located at $x=d/2$ on the top prestressing strand embedded in the sleeper during the testing procedure; a large increase at strain between 1100 and 1150 seconds was reported to likely be because of cracking along the top of the sleeper \citep{Butler}. The Gaussian process model is trained on (i.e. estimating $\widehat{\mu}_f$) the observed curvature data corresponding to the first 100 seconds of FBG strain measurements (5000 data points recorded at 50Hz) using Algorithm \ref{alg:general}. The tuning procedure outlined in Sec. \ref{sec:adaptivenu} is used to specify the value of $N_u$ used for the training of the Gaussian process model and this results in $N_{u}=5$. $EI$ was estimated as 9.07$\times 10^{11}$, and $k$ was estimated as 219.41. The parameter for the covariance kernels, $\sigma^2$, was estimated as 24.6. These estimates make up the set $\widehat{\theta}$.

The log marginal likelihood $\log p(Y^{(i)}|\widehat{\theta})$ is computed for each 1000-wide batch of curvature data points for the entire experiment after the training period (refer to Figure \ref{figure:log_likelihood_sleeper_experiment}). The likelihood very slightly decreases over time, but sharply drops between 1100 seconds and 1120 seconds (55000'th and 56000'th data points) signifying a change in the structural state of the system. This is inline with the change-point observed in \citep{Butler} and shown by the strain measurements in Figure \ref{figure:strain_sleeper_experiment}, and so the proposed modelling method works as expected alongside this experimental setup. It uses the drop-off in log marginal likelihood as a precursor for a visibly detected structural change (e.g. concrete cracking).

\section{Conclusions}
\label{sec:conclusion}

This paper has presented an important contribution to the structural identification and health monitoring community. The contribution takes the form of a multi-output data-centric engineering Gaussian process model that fuses together information from an analytical physics-based model for the deflection and observed data for the curvature of an instrumented railway sleeper supported on compacted ballast. This fusion is implemented on the level of the geometric relationship between these different physical quantities. The multi-output Gaussian process model provides aposteriori estimates to the curvature of the sleeper, conditioned on both the observed data and physics-based model. This approach allows the physics-based model to guide the inference from the data in unmeasured regions of the domain.
The resulting methodology also produces estimates to parameters within the physics-based model, such as flexural rigidity and ballast stiffness. These are used for the detection of system changes; underlying changes in the parameters of the physics-based model can indicate potential structural damage.

This work builds upon a surge of recent work within the model updating community \citep{Vigliotti, Schommer, Alkayem}. The estimation in the material properties of beams, such as flexural rigidity, has been a major area of interest in such literature.
As demonstrated in this study, these methods even have the potential to shed light on material properties (and therefore any changes) in components of physics-based models that are not measured by instrumented sensors such as the ballast supporting railway sleepers. A limitation of the proposed methodology is the assumption that the flexural rigidity utilised in the physics-based model is constant over the sleeper. A future aim for this line of research is to modify the methodology to relax this assumption, perhaps by assuming a functional form for the flexural rigidity (e.g. piecewise linear) and estimating the piecewise coefficients. The physics-based model used can also be generalised to something more sophisticated (e.g. FEM \citep{Zhou}) in order to model a larger structure, such as an entire railway bridge rather than only the sleeper component. In addition to this, aposteriori estimates of sleeper response obtained through the presented methodology could inform the optimal placement of sensors \citep{Guratzsch} and the detection of faulty sensors \citep{Hernandez}, possibly through active learning. An attractive side effect of the proposed technique is the potential to use it as a method for signal-level sensor fusion \citep{Soman, Hall} (the combination of data from multiple sensor networks). For such an implementation, the simulations from the analytical physics-based model in this work would be exchanged for observed data from another sensor network measuring vertical deflection, e.g. deflection transducers \citep{Rodrigues}.

This paper demonstrates the proposed methodology using simulated and experimental curvature data. The experimental data is the product of a recent full-scale laboratory testing procedure for a railway sleeper supported on compacted ballast, outlined in \cite{Xu}. The testing procedure imparts forcing onto a sleeper instrumented with fibre optic based strain sensors until damage occurs. The proposed model in this paper detects such structural change via evaluating the marginal likelihood with future sensor data. Another important development made in this work is the process of balancing the information obtained from the observed data and the physics-based model used in training the multi-output Gaussian process model. This is done in order to improve the predictive performance of a test set of observed data assumed to be from unmeasured regions of the domain.
This type of methodology is aimed at inspiring future work in the development and evaluation of data-centric engineering models.

\section*{Acknowledgements}
This work was supported by The Alan Turing Institute under the EPSRC grant EP/N510129/1 and the Turing-Lloyd’s Register Foundation Programme for Data-Centric Engineering. The authors would also like to acknowledge EPSRC and Innovate UK (grant no. 920035) for funding this research through the Centre for Smart Infrastructure and Construction (CSIC) Innovation and Knowledge Centre. Research related to installation of the sensor system was carried out under EPSRC grant no. EP/N021614. Mark Girolami is supported by a Royal Academy of Engineering Research Chair in Data Centric Engineering.

\appendix

\bibliographystyle{elsarticle-harv} 
\bibliography{refs}

\begin{thebibliography}{47}
\expandafter\ifx\csname natexlab\endcsname\relax\def\natexlab#1{#1}\fi
\providecommand{\url}[1]{\texttt{#1}}
\providecommand{\href}[2]{#2}
\providecommand{\path}[1]{#1}
\providecommand{\DOIprefix}{doi:}
\providecommand{\ArXivprefix}{arXiv:}
\providecommand{\URLprefix}{URL: }
\providecommand{\Pubmedprefix}{pmid:}
\providecommand{\doi}[1]{\href{http://dx.doi.org/#1}{\path{#1}}}
\providecommand{\Pubmed}[1]{\href{pmid:#1}{\path{#1}}}
\providecommand{\bibinfo}[2]{#2}
\ifx\xfnm\relax \def\xfnm[#1]{\unskip,\space#1}\fi
\bibitem[{Aktan and Brownjohn(2013)}]{Aktan}
\bibinfo{author}{Aktan, A.E.}, \bibinfo{author}{Brownjohn, J.M.W.},
  \bibinfo{year}{2013}.
\newblock \bibinfo{title}{Structural identification: Opportunities and
  challenges}.
\newblock \bibinfo{journal}{Journal of Structural Engineering}
  \bibinfo{volume}{139}, \bibinfo{pages}{1639--1647}.
\bibitem[{Alkayem et~al.(2017)Alkayem, Cao, Zhang, Bayat and Su}]{Alkayem}
\bibinfo{author}{Alkayem, N.F.}, \bibinfo{author}{Cao, M.},
  \bibinfo{author}{Zhang, Y.}, \bibinfo{author}{Bayat, M.},
  \bibinfo{author}{Su, Z.}, \bibinfo{year}{2017}.
\newblock \bibinfo{title}{Structural damage detection using finite element
  model updating with evolutionary algorithms: a survey}.
\newblock \bibinfo{journal}{Neural Computing and Applications} ,
  \bibinfo{pages}{1--23}.
\bibitem[{Blondeel et~al.(2018)Blondeel, Robbe, Lombaert and
  Vandewalle}]{Blondeel}
\bibinfo{author}{Blondeel, P.}, \bibinfo{author}{Robbe, P.},
  \bibinfo{author}{Lombaert, G.}, \bibinfo{author}{Vandewalle, S.},
  \bibinfo{year}{2018}.
\newblock \bibinfo{title}{Multilevel monte carlo for uncertainty quantification
  in structural engineering}.
\newblock \bibinfo{journal}{arXiv preprint arXiv:1808.10680} .
\bibitem[{Brooks and Morgan(1995)}]{Brooks}
\bibinfo{author}{Brooks, S.P.}, \bibinfo{author}{Morgan, B.J.T.},
  \bibinfo{year}{1995}.
\newblock \bibinfo{title}{Optimization using simulated annealing}.
\newblock \bibinfo{journal}{The Statistician} , \bibinfo{pages}{241--257}.
\bibitem[{{\c{C}}atba{\c{s}} et~al.(2013){\c{C}}atba{\c{s}}, Kijewski-Correa
  and Aktan}]{ccatbacs2013structural}
\bibinfo{author}{{\c{C}}atba{\c{s}}, F.N.}, \bibinfo{author}{Kijewski-Correa,
  T.}, \bibinfo{author}{Aktan, A.E.}, \bibinfo{year}{2013}.
\newblock \bibinfo{title}{Structural identification of constructed systems:
  approaches, methods, and technologies for effective practice of st-id},
  \bibinfo{organization}{American Society of Civil Engineers}.
\bibitem[{Cawley(2018)}]{Cawley}
\bibinfo{author}{Cawley, P.}, \bibinfo{year}{2018}.
\newblock \bibinfo{title}{Structural health monitoring: Closing the gap between
  research and industrial deployment}.
\newblock \bibinfo{journal}{Structural Health Monitoring}
  \DOIprefix\doi{10.1177/1475921717750047}.
\bibitem[{Dawari and Vesmawala(2013)}]{Dawari}
\bibinfo{author}{Dawari, V.B.}, \bibinfo{author}{Vesmawala, G.R.},
  \bibinfo{year}{2013}.
\newblock \bibinfo{title}{Structural damage identification using modal
  curvature differences}.
\newblock \bibinfo{journal}{IOSR J Mech Civ Eng} \bibinfo{volume}{4},
  \bibinfo{pages}{33--38}.
\bibitem[{Doebling et~al.(1996)Doebling, Farrar, Prime and Shevitz}]{Doebling}
\bibinfo{author}{Doebling, S.W.}, \bibinfo{author}{Farrar, C.R.},
  \bibinfo{author}{Prime, N.B.}, \bibinfo{author}{Shevitz, D.W.},
  \bibinfo{year}{1996}.
\newblock \bibinfo{title}{Damage identification and health monitoring of
  structural and mechanical systems from changes in their vibration
  characteristics: a literature review} .
\bibitem[{Farrar and Worden(2007)}]{farrar2007introduction}
\bibinfo{author}{Farrar, C.R.}, \bibinfo{author}{Worden, K.},
  \bibinfo{year}{2007}.
\newblock \bibinfo{title}{An introduction to structural health monitoring}.
\newblock \bibinfo{journal}{Philosophical Transactions of the Royal Society of
  London A: Mathematical, Physical and Engineering Sciences}
  \bibinfo{volume}{365}, \bibinfo{pages}{303--315}.
\bibitem[{Farrar and Worden(2012)}]{farrar2012structural}
\bibinfo{author}{Farrar, C.R.}, \bibinfo{author}{Worden, K.},
  \bibinfo{year}{2012}.
\newblock \bibinfo{title}{Structural health monitoring: a machine learning
  perspective}.
\newblock \bibinfo{publisher}{John Wiley \& Sons}.
\bibitem[{Fawcett(2006)}]{Fawcett}
\bibinfo{author}{Fawcett, T.}, \bibinfo{year}{2006}.
\newblock \bibinfo{title}{An introduction to roc analysis}.
\newblock \bibinfo{journal}{Pattern recognition letters} \bibinfo{volume}{27},
  \bibinfo{pages}{861--874}.
\bibitem[{Friswell and Penny(2002)}]{Friswell}
\bibinfo{author}{Friswell, M.I.}, \bibinfo{author}{Penny, J.E.T.},
  \bibinfo{year}{2002}.
\newblock \bibinfo{title}{Crack modeling for structural health monitoring}.
\newblock \bibinfo{journal}{Structural health monitoring} \bibinfo{volume}{1},
  \bibinfo{pages}{139--148}.
\bibitem[{Fuentes et~al.()Fuentes, Cross, Halfpenny, Barthorpe and
  Worden}]{Fuentes}
\bibinfo{author}{Fuentes, R.}, \bibinfo{author}{Cross, E.J.},
  \bibinfo{author}{Halfpenny, A.}, \bibinfo{author}{Barthorpe, R.J.},
  \bibinfo{author}{Worden, K.}, .
\newblock \bibinfo{title}{Autoregressive gaussian processes for structural
  damage detection} .
\bibitem[{Grafe()}]{Grafe}
\bibinfo{author}{Grafe, H.}, .
\newblock \bibinfo{title}{Model updating of large structural dynamics models
  using measured response functions}.
\newblock Ph.D. thesis.
\bibitem[{Gregory et~al.(2018)Gregory, Lau and Butler}]{Gregory}
\bibinfo{author}{Gregory, A.}, \bibinfo{author}{Lau, F.},
  \bibinfo{author}{Butler, L.}, \bibinfo{year}{2018}.
\newblock \bibinfo{title}{A quantile-based approach to modelling recovery time
  in structural health monitoring}.
\newblock \bibinfo{journal}{arXiv preprint arXiv:1803.08444} .
\bibitem[{Guratzsch and Mahadevan(2010)}]{Guratzsch}
\bibinfo{author}{Guratzsch, R.F.}, \bibinfo{author}{Mahadevan, S.},
  \bibinfo{year}{2010}.
\newblock \bibinfo{title}{Structural health monitoring sensor placement
  optimization under uncertainty}.
\newblock \bibinfo{journal}{AIAA journal} \bibinfo{volume}{48},
  \bibinfo{pages}{1281--1289}.
\bibitem[{Hall and Llinas(1997)}]{Hall}
\bibinfo{author}{Hall, D.L.}, \bibinfo{author}{Llinas, J.},
  \bibinfo{year}{1997}.
\newblock \bibinfo{title}{An introduction to multisensor data fusion}.
\newblock \bibinfo{journal}{Proceedings of the IEEE} \bibinfo{volume}{85},
  \bibinfo{pages}{6--23}.
\bibitem[{Hernandez-Garcia and Masri(2014)}]{Hernandez}
\bibinfo{author}{Hernandez-Garcia, M.R.}, \bibinfo{author}{Masri, S.F.},
  \bibinfo{year}{2014}.
\newblock \bibinfo{title}{Application of statistical monitoring using
  latent-variable techniques for detection of faults in sensor networks}.
\newblock \bibinfo{journal}{Journal of Intelligent Material Systems and
  Structures} \bibinfo{volume}{25}, \bibinfo{pages}{121--136}.
\bibitem[{Het{\'e}nyi(1971)}]{Hetenyi}
\bibinfo{author}{Het{\'e}nyi, M.}, \bibinfo{year}{1971}.
\newblock \bibinfo{title}{Beams on elastic foundation: theory with applications
  in the fields of civil and mechanical engineering}.
\newblock \bibinfo{publisher}{University of Michigan}.
\bibitem[{Kennedy and O'Hagan(2001)}]{OHagan}
\bibinfo{author}{Kennedy, M.C.}, \bibinfo{author}{O'Hagan, A.},
  \bibinfo{year}{2001}.
\newblock \bibinfo{title}{Bayesian calibration of computer models}.
\newblock \bibinfo{journal}{Journal of the Royal Statistical Society: Series B
  (Statistical Methodology)} \bibinfo{volume}{63}, \bibinfo{pages}{425--464}.
\bibitem[{Kreuzer(2006)}]{Kreuzer}
\bibinfo{author}{Kreuzer, M.}, \bibinfo{year}{2006}.
\newblock \bibinfo{title}{Strain measurement with fiber {B}ragg grating
  sensors} .
\bibitem[{Lau et~al.(2018a)Lau, Adams, Girolami, Butler and Elshafie}]{LauProb}
\bibinfo{author}{Lau, F.D.H.}, \bibinfo{author}{Adams, N.M.},
  \bibinfo{author}{Girolami, M.}, \bibinfo{author}{Butler, L.J.},
  \bibinfo{author}{Elshafie, M.Z.E.B.}, \bibinfo{year}{2018}a.
\newblock \bibinfo{title}{The role of statistics in data-centric engineering}.
\newblock \bibinfo{journal}{Statistics \& Probability Letters}
  \bibinfo{volume}{136}, \bibinfo{pages}{58--62}.
\bibitem[{Lau et~al.(2018b)Lau, Butler, Adams, Elshafie and Girolami}]{Lau}
\bibinfo{author}{Lau, F.D.H.}, \bibinfo{author}{Butler, L.},
  \bibinfo{author}{Adams, N.}, \bibinfo{author}{Elshafie, M.},
  \bibinfo{author}{Girolami, M.}, \bibinfo{year}{2018}b.
\newblock \bibinfo{title}{Real-time statistical modelling of data generated
  from self-sensing bridges}, in: \bibinfo{booktitle}{Proceedings of the
  Institution of Civil Engineers - Smart Infrastructure and Construction}.
\bibitem[{Liu and M{\"u}ller(2004)}]{Liu}
\bibinfo{author}{Liu, X.}, \bibinfo{author}{M{\"u}ller, H.G.},
  \bibinfo{year}{2004}.
\newblock \bibinfo{title}{Functional convex averaging and synchronization for
  time-warped random curves}.
\newblock \bibinfo{journal}{Journal of the American Statistical Association}
  \bibinfo{volume}{99}, \bibinfo{pages}{687--699}.
\bibitem[{Neves et~al.(2017)Neves, Gonz{\'a}lez, Leander and Karoumi}]{Neves}
\bibinfo{author}{Neves, A.C.}, \bibinfo{author}{Gonz{\'a}lez, I.},
  \bibinfo{author}{Leander, J.}, \bibinfo{author}{Karoumi, R.},
  \bibinfo{year}{2017}.
\newblock \bibinfo{title}{Structural health monitoring of bridges: a model-free
  ann-based approach to damage detection}.
\newblock \bibinfo{journal}{Journal of Civil Structural Health Monitoring}
  \bibinfo{volume}{7}, \bibinfo{pages}{689--702}.
\bibitem[{Raissi et~al.(2017)Raissi, Perdikaris and Karniadakis}]{Raissi}
\bibinfo{author}{Raissi, M.}, \bibinfo{author}{Perdikaris, P.},
  \bibinfo{author}{Karniadakis, G.E.}, \bibinfo{year}{2017}.
\newblock \bibinfo{title}{Machine learning of linear differential equations
  using gaussian processes}.
\newblock \bibinfo{journal}{Journal of Computational Physics}
  \bibinfo{volume}{348}, \bibinfo{pages}{683--693}.
\bibitem[{Rasmussen(2004)}]{Rasmussen}
\bibinfo{author}{Rasmussen, C.E.}, \bibinfo{year}{2004}.
\newblock \bibinfo{title}{Gaussian processes in machine learning}, in:
  \bibinfo{booktitle}{Advanced lectures on machine learning}.
  \bibinfo{publisher}{Springer}, pp. \bibinfo{pages}{63--71}.
\bibitem[{Robert(2014)}]{Robert}
\bibinfo{author}{Robert, C.}, \bibinfo{year}{2014}.
\newblock \bibinfo{title}{Machine learning, a probabilistic perspective}.
\newblock \bibinfo{publisher}{Taylor \& Francis}.
\bibitem[{Rocchetta et~al.(2018)Rocchetta, Broggi, Huchet and
  Patelli}]{Rocchetta}
\bibinfo{author}{Rocchetta, R.}, \bibinfo{author}{Broggi, M.},
  \bibinfo{author}{Huchet, Q.}, \bibinfo{author}{Patelli, E.},
  \bibinfo{year}{2018}.
\newblock \bibinfo{title}{On-line bayesian model updating for structural health
  monitoring}.
\newblock \bibinfo{journal}{Mechanical Systems and Signal Processing}
  \bibinfo{volume}{103}, \bibinfo{pages}{174--195}.
\bibitem[{Rodrigues et~al.(2011)Rodrigues, F{\'e}lix and Figueiras}]{Rodrigues}
\bibinfo{author}{Rodrigues, C.}, \bibinfo{author}{F{\'e}lix, C.},
  \bibinfo{author}{Figueiras, J.}, \bibinfo{year}{2011}.
\newblock \bibinfo{title}{Fiber-optic-based displacement transducer to measure
  bridge deflections}.
\newblock \bibinfo{journal}{Structural Health Monitoring} \bibinfo{volume}{10},
  \bibinfo{pages}{147--156}.
\bibitem[{Sandhu et~al.(2018)Sandhu, Reinarz and Dodwell}]{Dodwell}
\bibinfo{author}{Sandhu, A.}, \bibinfo{author}{Reinarz, A.},
  \bibinfo{author}{Dodwell, T.}, \bibinfo{year}{2018}.
\newblock \bibinfo{title}{A bayesian framework for assessing the strength
  distribution of composite structures with random defects}.
\newblock \bibinfo{journal}{arXiv preprint arXiv:1804.07549} .
\bibitem[{Sazonov et~al.(2004)Sazonov, Janoyan and Jha}]{Sazonov}
\bibinfo{author}{Sazonov, E.}, \bibinfo{author}{Janoyan, K.},
  \bibinfo{author}{Jha, R.}, \bibinfo{year}{2004}.
\newblock \bibinfo{title}{Wireless intelligent sensor network for autonomous
  structural health monitoring}, in: \bibinfo{booktitle}{Smart Structures and
  Materials 2004: Smart Sensor Technology and Measurement Systems},
  \bibinfo{organization}{International Society for Optics and Photonics}. pp.
  \bibinfo{pages}{305--315}.
\bibitem[{Schommer et~al.(2017)Schommer, Nguyen, Maas and
  Z{\"u}rbes}]{Schommer}
\bibinfo{author}{Schommer, S.}, \bibinfo{author}{Nguyen, V.H.},
  \bibinfo{author}{Maas, S.}, \bibinfo{author}{Z{\"u}rbes, A.},
  \bibinfo{year}{2017}.
\newblock \bibinfo{title}{Model updating for structural health monitoring using
  static and dynamic measurements}.
\newblock \bibinfo{journal}{Procedia engineering} \bibinfo{volume}{199},
  \bibinfo{pages}{2146--2153}.
\bibitem[{Seo et~al.(2000)Seo, Wallat, Graepel and Obermayer}]{Seo}
\bibinfo{author}{Seo, S.}, \bibinfo{author}{Wallat, M.},
  \bibinfo{author}{Graepel, T.}, \bibinfo{author}{Obermayer, K.},
  \bibinfo{year}{2000}.
\newblock \bibinfo{title}{Gaussian process regression: Active data selection
  and test point rejection}, in: \bibinfo{booktitle}{Mustererkennung 2000}.
  \bibinfo{publisher}{Springer}, pp. \bibinfo{pages}{27--34}.
\bibitem[{Sinha et~al.(2002)Sinha, Friswell and Edwards}]{Sinha}
\bibinfo{author}{Sinha, J.}, \bibinfo{author}{Friswell, M.I.},
  \bibinfo{author}{Edwards, S.}, \bibinfo{year}{2002}.
\newblock \bibinfo{title}{Simplified models for the location of cracks in beam
  structures using measured vibration data}.
\newblock \bibinfo{journal}{Journal of Sound and vibration}
  \bibinfo{volume}{251}, \bibinfo{pages}{13--38}.
\bibitem[{Soman et~al.(2018)Soman, Kyriakides, Onoufriou and
  Ostachowicz}]{Soman}
\bibinfo{author}{Soman, R.}, \bibinfo{author}{Kyriakides, M.},
  \bibinfo{author}{Onoufriou, T.}, \bibinfo{author}{Ostachowicz, W.},
  \bibinfo{year}{2018}.
\newblock \bibinfo{title}{Numerical evaluation of multi-metric data fusion
  based structural health monitoring of long span bridge structures}.
\newblock \bibinfo{journal}{Structure and Infrastructure Engineering}
  \bibinfo{volume}{14}, \bibinfo{pages}{673--684}.
\bibitem[{Sun et~al.(2010)Sun, Staszewski and Swamy}]{Sun}
\bibinfo{author}{Sun, M.}, \bibinfo{author}{Staszewski, W.J.},
  \bibinfo{author}{Swamy, R.N.}, \bibinfo{year}{2010}.
\newblock \bibinfo{title}{Smart sensing technologies for structural health
  monitoring of civil engineering structures}.
\newblock \bibinfo{journal}{Advances in Civil Engineering}
  \bibinfo{volume}{2010}.
\bibitem[{Teimouri et~al.(2017)Teimouri, Milani, Loeppky and
  Seethaler}]{Teimouri}
\bibinfo{author}{Teimouri, H.}, \bibinfo{author}{Milani, A.S.},
  \bibinfo{author}{Loeppky, J.}, \bibinfo{author}{Seethaler, R.},
  \bibinfo{year}{2017}.
\newblock \bibinfo{title}{A gaussian process--based approach to cope with
  uncertainty in structural health monitoring}.
\newblock \bibinfo{journal}{Structural Health Monitoring} \bibinfo{volume}{16},
  \bibinfo{pages}{174--184}.
\bibitem[{Tran et~al.(2017)Tran, Hoang, Foret, Duhamel, Messad and
  Loaec}]{Tran}
\bibinfo{author}{Tran, L.H.}, \bibinfo{author}{Hoang, T.},
  \bibinfo{author}{Foret, G.}, \bibinfo{author}{Duhamel, D.},
  \bibinfo{author}{Messad, S.}, \bibinfo{author}{Loaec, A.},
  \bibinfo{year}{2017}.
\newblock \bibinfo{title}{Analytical model of the dynamics of railway sleeper},
  in: \bibinfo{booktitle}{6 th ECCOMAS Thematic Conference on Computational
  Methods in Structural Dynamics and Earthquake Engineering}, pp.
  \bibinfo{pages}{15--17}.
\bibitem[{Vigliotti et~al.(2018)Vigliotti, Cs{\'a}nyi and
  Deshpande}]{Vigliotti}
\bibinfo{author}{Vigliotti, A.}, \bibinfo{author}{Cs{\'a}nyi, G.},
  \bibinfo{author}{Deshpande, V.S.}, \bibinfo{year}{2018}.
\newblock \bibinfo{title}{Bayesian inference of the spatial distributions of
  material properties}.
\newblock \bibinfo{journal}{Journal of the Mechanics and Physics of Solids}
  \bibinfo{volume}{118}, \bibinfo{pages}{74--97}.
\bibitem[{Wehbi and Musgrave(2017)}]{PWI}
\bibinfo{author}{Wehbi, M.}, \bibinfo{author}{Musgrave, P.},
  \bibinfo{year}{2017}.
\newblock \bibinfo{title}{Optimisation of track stiffness on the {UK}
  railways}.
\newblock \bibinfo{journal}{Permanent Way Institute Journal}
  \bibinfo{volume}{135}.
\bibitem[{Worden and Cross(2018)}]{Worden}
\bibinfo{author}{Worden, K.}, \bibinfo{author}{Cross, E.J.},
  \bibinfo{year}{2018}.
\newblock \bibinfo{title}{On switching response surface models, with
  applications to the structural health monitoring of bridges}.
\newblock \bibinfo{journal}{Mechanical Systems and Signal Processing}
  \bibinfo{volume}{98}, \bibinfo{pages}{139--156}.
\bibitem[{Xu et~al.(2015)Xu, Ren and Wang}]{Xu}
\bibinfo{author}{Xu, H.}, \bibinfo{author}{Ren, W.X.}, \bibinfo{author}{Wang,
  Z.C.}, \bibinfo{year}{2015}.
\newblock \bibinfo{title}{Deflection estimation of bending beam structures
  using fiber bragg grating strain sensors}.
\newblock \bibinfo{journal}{Advances in Structural Engineering}
  \bibinfo{volume}{18}, \bibinfo{pages}{395--403}.
\bibitem[{Xu et~al.(2019)Xu, Butler and Elshafie}]{Butler}
\bibinfo{author}{Xu, J.}, \bibinfo{author}{Butler, L.},
  \bibinfo{author}{Elshafie, M.}, \bibinfo{year}{2019}.
\newblock \bibinfo{title}{Experimental and numerical investigation of the
  performance of self-sensing concrete sleepers}.
\newblock \bibinfo{journal}{Journal of Structural Health Monitoring.}
  \DOIprefix\doi{10.1177/1475921719834506}.
\bibitem[{Ye et~al.(2014)Ye, Su and Han}]{Ye}
\bibinfo{author}{Ye, X.W.}, \bibinfo{author}{Su, Y.H.}, \bibinfo{author}{Han,
  J.P.}, \bibinfo{year}{2014}.
\newblock \bibinfo{title}{Structural health monitoring of civil infrastructure
  using optical fiber sensing technology: A comprehensive review}.
\newblock \bibinfo{journal}{The Scientific World Journal}
  \bibinfo{volume}{2014}.
\bibitem[{Zeinali(2017)}]{Zeinali}
\bibinfo{author}{Zeinali, Y.}, \bibinfo{year}{2017}.
\newblock \bibinfo{title}{Framework for flexural rigidity estimation in
  euler-bernoulli beams using deformation influence lines}.
\newblock \bibinfo{journal}{Infrastructures} \bibinfo{volume}{2},
  \bibinfo{pages}{23}.
\bibitem[{Zhou and Tang(2018)}]{Zhou}
\bibinfo{author}{Zhou, K.}, \bibinfo{author}{Tang, J.}, \bibinfo{year}{2018}.
\newblock \bibinfo{title}{Uncertainty quantification in structural dynamic
  analysis using two-level gaussian processes and bayesian inference}.
\newblock \bibinfo{journal}{Journal of Sound and Vibration}
  \bibinfo{volume}{412}, \bibinfo{pages}{95--115}.

\end{thebibliography}





\end{document}